\documentclass[aps,pra,twocolumn,showpacs,10pt,amsmath,footinbib]{revtex4-1}
\usepackage{graphicx}
\usepackage[usenames,dvipsnames]{xcolor}
\usepackage{braket}
\usepackage{array}
\usepackage{amsmath}

\newcommand{\pa}{\partial}

\def\gappeq{\mathrel{ \rlap{\raise.5ex\hbox{$>$}}
                      {\lower.5ex\hbox{$\sim$}} } }
\def\lappeq{\mathrel{ \rlap{\raise.5ex\hbox{$<$}}
                      {\lower.5ex\hbox{$\sim$}} } }

                      \newcommand{\del}[1]{\textcolor{red}{}}
                      
\begin{document}

\title{Spin-up of a superfluid vortex lattice driven by rough boundaries}
%Rotating buckets of superfluid revisited: role of surface roughness}

\author{N. A. Keepfer} \email{n.a.l.keepfer1@ncl.ac.uk}
\affiliation{Joint Quantum Centre (JQC) Durham--Newcastle, 
School of Mathematics, Statistics and Physics, Newcastle University, 
Newcastle upon Tyne, NE1 7RU, United Kingdom, and
INO-CNR BEC Center, Dipartimento di Fisica, Università di Trento, 
via Sommarive 14, I-38123 Trento, Italy}

\author{G. W. Stagg} 
\affiliation{Joint Quantum Centre (JQC) Durham--Newcastle, School of Mathematics, Statistics and Physics, Newcastle University, Newcastle upon Tyne, NE1 7RU, United Kingdom}

\author{L. Galantucci} 
\affiliation{Joint Quantum Centre (JQC) Durham--Newcastle, School of Mathematics, Statistics and Physics, Newcastle University, Newcastle upon Tyne, NE1 7RU, United Kingdom}

\author{C. F. Barenghi} 
\affiliation{Joint Quantum Centre (JQC) Durham--Newcastle, School of Mathematics, Statistics and Physics, Newcastle University, Newcastle upon Tyne, NE1 7RU, United Kingdom}

\author{N. G. Parker} 
\affiliation{Joint Quantum Centre (JQC) Durham--Newcastle, School of Mathematics, Statistics and Physics, Newcastle University, Newcastle upon Tyne, NE1 7RU, United Kingdom}

\date{\today}

\begin{abstract}
We study numerically the formation of a vortex lattice inside a rotating bucket 
containing superfluid helium, paying attention to an important feature which
is practically unavoidable in all experiments: the microscopic roughness of the bucket's 
surface. 
We model this using the Gross-Pitaevskii equation for a weakly-interacting 
Bose gas, a model which is idealised when applied to superfluid helium but
captures the key physics of the vortex dynamics which we are interested in.  
We find that the vortex lattice arises from the interaction
and reconnections of nucleated U-shaped vortex lines, which merge and align
along the axis of rotation. We quantify the effects which the surface roughness
and remanent vortex lines play in this process. 
\end{abstract}

\maketitle

\section{Introduction}
\label{section1}

Superfluids are extraordinary fluids characterised by the absence of 
viscosity.
They are irrotational everywhere except at vortex lines
whose circulation is quantised in units of $\kappa=h/m$, where 
$h$ is Planck's constant and $m$ is the 
mass of the boson which composes the fluid \cite{Annett2004,Barenghi2016}.
First discovered and studied in liquid helium-4 and, decades later,
in helium-3, superfluidity has since been observed in ultracold gases and 
photonic systems. The constraint of quantised
vorticity is a consequence of quantum mechanics - vorticity can only arise as $2 \pi$ topological defects of the macroscopic single-particle wavefunction of the quantum many-body system.  These defects manifest as vortex lines through the fluid.  As well as possessing a circulating flow, the vortex lines have a core of depleted density about their axis, out to a core radius $a_0$ which is of the order of the superfluid healing length. In helium-4 and helium-3 the vortex core size is around $10^{-10}$m and $10^{-8}$m, respectively.

The textbook paradigm of superfluidity is a cylindrical
bucket of superfluid helium
rotating at constant angular frequency $\Omega$. 
Classical solid-body 
rotation is forbidden by the irrotational nature of the superfluid.  At  sufficiently small values of
$\Omega$, the fluid remains quiescent. However, if $\Omega$
in increased past a critical
value $\Omega_{\rm c}$, the presence of a vortex line
is energetically favourable.  Using hydrodynamic arguments and up to 
a logarithmic correction, it is estimated \cite{Barenghi2016} that this
critical angular frequency is

\begin{equation}
\Omega_{\rm c} =  \frac{\hbar}{mR^2} \ln \left( \frac{R}{a_0} \right),
\label{eqn:omega_c}
\end{equation}

\noindent
where $m$ is the mass of a helium atom and $R$ is the radius of the bucket.  At larger values of
$\Omega$, two vortices become favourable, and so on. 
For $\Omega \gg \Omega_{\rm c}$ the stationary state of the fluid is 
the famed vortex lattice, an array of vortex lines aligned along the
axis of rotation with areal density
\begin{equation}
n_{\rm v}=\frac{2\Omega}{\kappa},
\label{eq:Feynman}
\end{equation}
known as Feynman's rule. The vortex lattice was  
first imaged in superfluid helium by Packard {\it et al} \cite{Yarmchuk1979} 
and more recently by Bewley {\it et al} \cite{Bewley2006}.  
The lattice has also been observed in ultracold gaseous superfluids
trapped by smooth confining 
potentials \cite{Madison2000,AboShaeer2001}. (At much higher rotation frequencies where centrifugal effects dominate, a giant or macroscopic vortex carrying many quanta of circulation can become formed in both superfluid helium \cite{Tsakadze1964,Josserand2004} and gaseous superfluids \cite{Kasamatsu2002,Fischer2003,Kavoulakis2003,Engels2003}; however, this regime is outside the scope of this work.)

The process in which the vortices enter the superfluid in the first
place is called vortex nucleation. 
Being associated with a $2\pi$ phase singularity of the macroscopic 
wavefunction, a vortex line is topologically protected.
Thus, starting from some initially vortex-free state, vortices must enter
the superfluid from the boundary. It is believed that
vortex lines are nucleated
either {\it intrinsically} by the flow of the superfluid past the 
microscopic roughness of the bucket wall (overcoming a critical velocity)
or {\it extrinsically} by stretching some pre-existing vortex lines 
called ``remanent vortices" which,
under suitably conditions, can spool additional 
vortices \cite{Schwarz-mill}.  Remanent vortices are thought 
to arise when cooling the helium sample through the superfluid transition, 
and can be avoided by using careful
experimental protocols \cite{Yano-2007}.  

Individual vortex nucleation in a rotating bucket,
either intrinsic or extrinsic, has never been visualised in detail. 
Experimentally, it remains challenging to image the flow in 
the vicinity of a boundary, despite progress in flow visualisation 
in the bulk \cite{Bewley2006,Zmeev2015b,Duda2015}, more so because
the microscopic scale of the vortices themselves.  
Theoretically, the nucleation problem has been addressed using
energy arguments \cite{Fetter1966,StaufferFetter1968} with no insight
in the dynamics. With few exceptions \cite{Stagg2017},
the effect of microscopic boundary roughness on the vortex nucleation has not been studied. 
A related and better understood nucleation  process
takes place when an ion bubble is driven in liquid helium by an applied 
electric field; compared to the bucket, the nucleation is more 
controlled in terms of geometry (the shape of the bubble can be
determined theoretically)
and velocity (experimentally determined by time of flight measurements).
Vortex nucleation by the ion bubble has thus received much detailed 
experimental and theoretical attention
\cite{MuirheadVinenDonnelly1984,McClintockBowley1995,BerloffRoberts2000,
Winiecki2000,Villois2018} than nucleation by the walls of the bucket 
which contains the helium sample.

In this work we are not concerned with the vortex nucleation as such,
but rather with the intermediate state between the nucleation and
the final vortex lattice. This intermediate stage is still unexplored, 
but, given that the length scales and the time scales involved depend on
the vortex separation rather the vortex core size (i.e. they are
mesoscopic rather than microscopic), there is prospect of 
experimental visualisation in the near future.
The focus of attention is therefore not individual vortex dynamics at 
nucleation but the collective dynamics of many vortex lines in the 
presence of a boundary which is not smooth. For simplicity we consider 
the problem at sufficiently low temperature that the normal fluid does
not play an important role.

The traditional method to model the dynamics of superfluid vortices is
the Vortex Filament Method (VFM) \cite{Schwarz1988}, 
which models vortex lines as infinitesimally thin filaments 
interacting with themselves, their neighbours and the boundary 
(via suitable images).  However, this approach is not applicable
to our problem.
Firstly, if the boundary varies on atomic length scales comparable 
to the vortex core (which is likely to be the case for any metal or glass 
bucket containing liquid helium),
then the core lengthscale can no longer be ignored compared to
other relevant lengthscale, invalidating the 
assumptions behind the VFM.
Secondly, the implementation of the boundary
condition is cumbersome to set up and
not simple to change from one boundary shape to another; indeed, the VFM 
has been implemented for plane \cite{Schwarz1985}, 
semi-spherical \cite{Schwarz1985,Tsubota1993}, 
spherical \cite{Schwarz1974,Kivotides2006} and cylindrical
\cite{Hanninen2005,HanninenBaggaley2014} boundaries, but never
for irregular boundaries relevant to our problem. Thirdly, the VFM does
not describe vortex nucleation, but requires to initialise the calculation
with arbitrary seeding vortex lines.
An alternative approach is through the Gross-Pitaevskii equation (GPE) 
\cite{Pitaevskii,Barenghi2016}. This is a formal description 
of a dilute weakly-interacting gas of bosons, and is equivalent 
to a continuity equation and an Euler-like equation for an inviscid 
fluid (the modification being the presence
of a quantum pressure term).  While the GPE is an excellent quantitative 
description of Bose gas superfluids, it is limited to being a qualitative 
description of superfluid helium due to the stronger interactions
taking place in a liquid rather than in a gas.  
Nevertheless, its capability to describe the microscopic detail of 
superfluid dynamics - the finite-sized core, 
vortex interactions and reconnections, even the intrinsic nucleation - makes 
it a useful model to study superfluid flows at a boundary.  
An important feature is that the GPE can easily implement
irregular boundaries.
Indeed, recent GPE simulations have predicted the occurrence of a 
turbulent boundary layer when the superfluid flows past a locally
rough surface \cite{Stagg2017}:  above a critical imposed flow speed, 
vortices are nucleated from the surface features, interact and become 
entwined in a layer adjacent to the surface.  

Returning to the rotating bucket of superfluid helium, 
it is natural to ask if some kind of boundary layer may similarly 
form at the boundary of the rotating bucket
in the transient evolution to the vortex lattice. Whether disordered
or laminar, this layer will certainly involve vortex interactions.
It is in fact unlikely that the vortex lines which nucleate extend
from the top to the bottom of the bucket, as if the process were
essentially two-dimensional (2D). More likely, the first vortex lines
which nucleate are small, and become long only after a sequence of
interactions and reconnections.
To qualitatively explore these interactions, here we perform a series of 
numerical experiments, based on the GPE, of a superfluid being spun-up 
in a bucket whose walls are microscopically rough.  
These numerical experiments allows us to build a physical picture 
of how vorticity enters the superfluid and forms a vortex lattice, 
and of the role of remanent vortices, 
sharp intrusions, rotation rate, and dimensionality.  

The plan of the paper is the following.
In Section~\ref{section2} we introduce our model and details our of numerical
simulations.  In Section~\ref{section3} we present our main results for the
spin up of a quiescent superfluid. Section~\ref{section4} explores
the possibility that a single strong imperfection in the shape of
a protuberance, remanent vortex lines or dimensionality may affect the main results described
in Section~\ref{section3}. Finally, in Section~\ref{section5} 
we discuss and conclude our findings.

%Theoretical progress is challenging and to date has
%focussed on smooth and idealised surfaces. In principle, the
%superfluid boundary conditions are straightforward: the
%superfluid velocity component, which is perpendicular to
%the boundary, must vanish at the boundary, whereas the
%tangential component (in the absence of viscous stresses)
%can slip. For the latter reason, in superfluids we do not
%expect boundary layers typical of viscous flows.
%Implementing these superfluid boundary conditions, it
%was found [24,25] that one or more vortices sliding along
%a smooth surface can become deflected or trapped by small
%hemispherical bumps. Such bumps can also serve as
%nucleation sites for vortices; the local superfluid velocity
%is raised at the pole of the bump and more readily breaks the
%critical velocity for vortex nucleation [26]. Indeed, our
%recent simulations [27] have shown that, if the bump is
%elliptically shaped and elongated perpendicular to the
%imposed flow, the superfluid velocity v at the pole is
%enhanced, reducing the critical Mach number for vortex
%nucleation from v=c ∼ 1 to smaller values v=c ∼ ϵ−1 ≪ 1
%(where ϵ ≫ 1 is the ellipticity of the bump), and increasing
%the intrinsic vortex nucleation rate (for a given supercritical
%imposed flow). We expect, therefore, that microscopically
%small surface roughness may promote the
%nucleation of vortices at a surface. For preexisting vortex

\section{Model and method}
\label{section2}

\subsection{Gross-Pitaevskii equation}

We model the superfluid dynamics using the Gross-Pitaevskii equation. Within this model, the superfluid is parametrised by a mean-field complex
wavefunction $\Psi({\bf r},t)=\vert \Psi({\bf r},t)\vert e^{iS(\bf r,t)}$.  
The particle density follows as $n({\bf r},t)=|\Psi({\bf r},t)|^2$ and 
the fluid velocity as ${\bf v}({\bf r},t)=(\hbar/m)\nabla S({\bf r},t)$, 
where $\hbar=h/(2 \pi)$ and $S({\bf r},t)$ is the phase distribution 
of $\Psi$.  The dynamics of $\Psi({\bf r},t)$ follows the 
GPE \cite{Pitaevskii,Barenghi2016},

\begin{equation}
i \hbar \frac{\pa\Psi }{\pa t}= \hat{\mathcal{H}} \Psi, 
\label{eq:GPE}
\end{equation}

\noindent
with Hamiltonian operator,

\begin{equation}
\hat{\mathcal{H}} = 
-\frac{\hbar^2}{2m}\nabla^2
+V+g\left|\Psi \right|^{2}.
\end{equation}

\noindent
Here $m$ is the particle mass, $g$ $(>0)$ is a nonlinear coefficient 
describing the inter-particle interactions,
 and $V({\bf r},t)$ is the external potential acting on the fluid.  
Stationary solutions of the GPE satisfy 
$\hat{\mathcal{H}}\Psi= \mu_0 \Psi$, 
where $\mu_0$ is the chemical potential of the fluid.   

We make two physically-motivated modifications to the basic GPE above.  
Firstly, since the GPE conserves energy, we follow other works 
\cite{Choi1998,Tsubota2001} in introducing a phenomenological dissipation 
term into the GPE to model, at least in a qualitative way, 
the damping of excitations of the superfluid (for example, by their 
interaction with the normal fluid).
This is achieved by replacing the left-side of Eq. (\ref{eq:GPE}) 
with $(i-\gamma) \hbar \, \partial \Psi/\partial t$, where $\gamma$ 
specifies the strength of the dissipation.  
Although not as accurate the friction included within the
 VFM, this phenomenological dissipation
will help damp out the oscillations of the vortex lines (Kelvin waves),
which is the main effect of the friction which concerns us here.
 Secondly, given our rotating scenario, we work in the reference
frame rotating at constant angular frequency $\Omega$ about the $z$ axis; this is achieved by modifying the GPE Hamiltonian to $\hat{\mathcal{H}}-i \Omega L_z$, where $L_z$ is the angular momentum operator about $z$.   In Cartesian coordinates $L_z=i \hbar (y \partial_x-x \partial_y)$.

\subsection{Bucket set-up}

We consider the fluid to be confined within 
a cylindrical bucket of radius $R$ and height $H$. The axis of the cylinder
is the z-axis of rotation.  
The bucket is modelled through the potential $V({\bf r})$: in the interior 
of the bucket we set $V=0$ while at the boundary and beyond we set 
$V \gg \mu_0$. In the ground state, the fluid density has the bulk value 
$n_0$ in the centre of the bucket, while close to the bucket wall 
it heals to zero density over a length scale characterised by the healing 
length $\xi=\hbar/\sqrt{m n_0 g}$.  The healing length also characterises 
the size of the cores of vortices in the fluid.
Note that the chemical potential in the bulk is $\mu_0=n_0 g$.  
The speed of sound in the uniform systems is
$c=\sqrt{n_0 g /m}$.

It is clearly computationally impossible
to simulate the range of length scales which are realistic for a typical
experiment with liquid helium in the context of the GPE model.
The dimensions (radius and height) of typical buckets used 
in the experiments are of the order of the
centimetre, which is around eight orders of magnitude larger than the 
vortex core size in helium-4, $a_0=10^{-10}\rm m$ (in helium-3 the vortex
core is about 100 times larger).
Instead, in our numerical experiments we employ buckets whose scale 
is around 2 orders of magnitude larger than the vortex core size.
While this is clearly a vast scale reduction compared to real systems, 
the separation of scales between the vortices and the bucket size 
is sufficient to give us a qualitative insight into the dynamics
of the vortex lines.

\subsection{Surface roughness}

\begin{figure}
\centering
\includegraphics[width = 0.9\columnwidth]{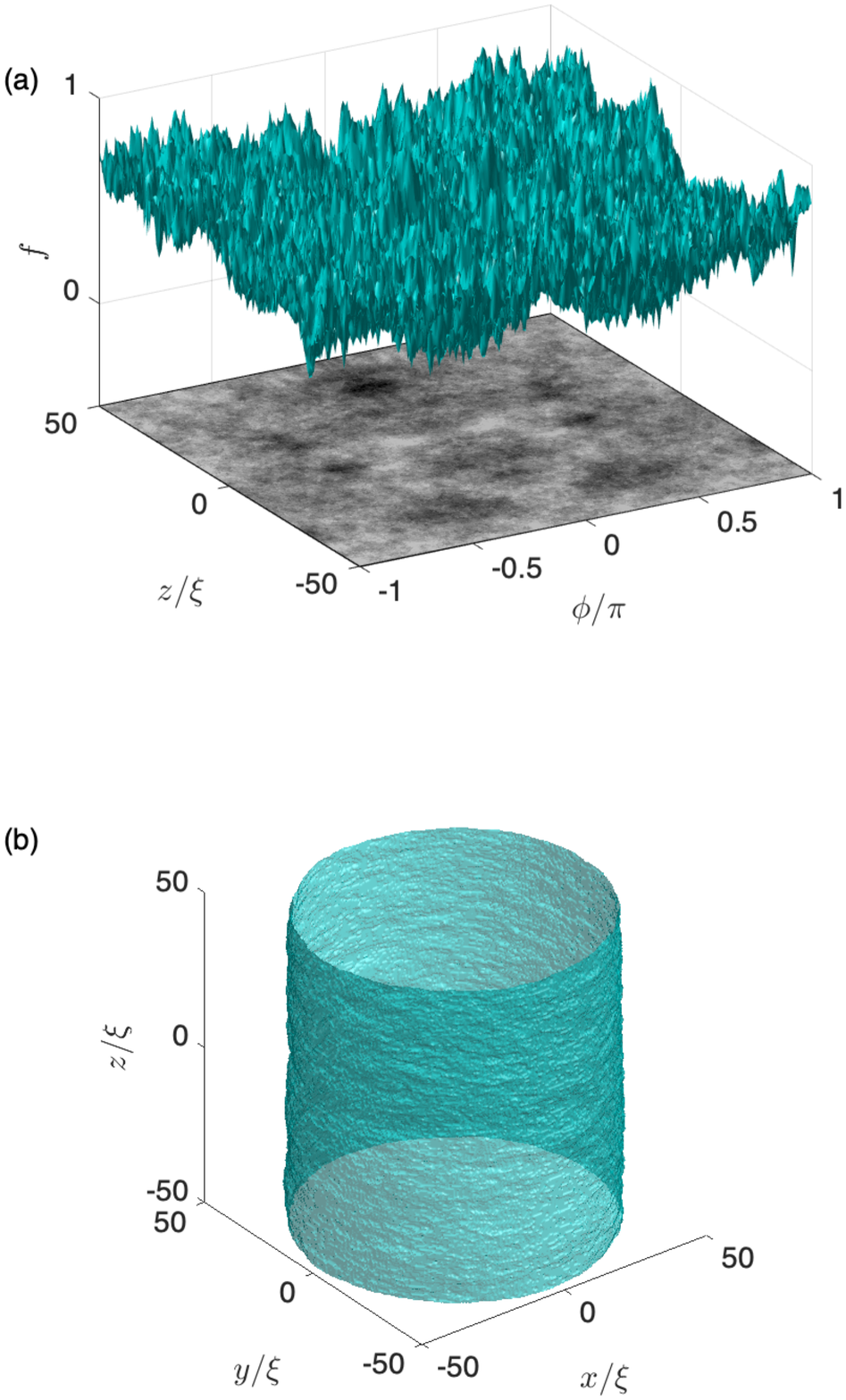}
\caption{(a) An example of the 2D fractional Brownian motion function $f$, 
normalised to the range [0,1], shown as a surface plot and a heatmap.
(b) The rough cylindrical boundary of our bucket
is formed by using the surface in (a) to modulate
the radius of the bucket boundary with an amplitude $a$. Here $a=0.1$. }
\label{fig1}
\end{figure}

To mimic the experimentally unavoidable surface roughness, 
we modify the azimuthal face of the bucket away from a perfect 
cylindrical shape using a noisy two-dimensional (2D) function.  
This function is numerically generated through a two-dimensional 
fractal Brownian motion \cite{MandelbrotVanNess}
with Hurst index of $0.3$, a parameter which describes the fractal dimension
of the surface \cite{Mandelbrot1985}.
The choice to model the roughness in this way is motivated by the well established fractal properties of real surfaces, including machined surfaces (of relevance to helium experiments), and the success of fractal brownian motion in modelling a wide variety of real rough surfaces \cite{Majumdar1990}.   
The function is normalised between $0$ and $1$, and is mirrored 
about its edge and recombined with itself in order to create periodicity 
across one dimension; a single realisation of the function is 
depicted in Fig.~\ref{fig1}(a).  The function is mapped onto 
the space of axial coordinate $z$ and azimuthal angle $\theta$, 
and used to modify the radius of the bucket according to the form,

\begin{equation}
r(z,\phi)=R(1-a f(z,\theta)),
\label{eq:rbucket}
\end{equation}

\noindent
where $R$ is the smooth bucket radius and $a$ is the 
(dimensionless) roughness parameter.
This numerical procedure
generates all of our rough 3D bucket shapes. 
%Note that the vertical and horizontal scales of Fig.~\ref{fig1} are
%not the same, therefore in the figure
%the rough surface may look more jagged than what it is in reality. 
By computing the local curvature of the surface roughness,
we find that the values of the average radius of curvature corresponding to
values $a=0.05$, $0.1$, $0.2$ and $0.3$ of the roughness parameter are
$10.4 \xi$, $5.2 \xi$, $2.6 \xi$ and $1.7\xi$ respectively
(small values of $a$ correspond to large radius of curvature, i.e. smoother
surface).
For simplicity, the top and bottom surfaces of the bucket are left
smooth. The reason is that, by providing the vortex lines with
pinning sites, any roughness on these surfaces 
will act essentially as an extra friction (an effect which already
qualitatively account for via the dissipation parameter $\gamma$) 
slowing down the final stage of cristallization of the vortex lattice.

\subsection{Simulation set-up}

The initial condition for $\Psi$ in all of our simulations is the 
non-rotating ground state solution, found by the method of imaginary 
time propagation of the GPE, supplemented with low-amplitude 
white noise to $\Psi$ (amplitude $0.001$) to break any symmetries 
artificially presented in the initial condition.   
We then impose a constant rotation on the system for $t>0$, 
with fixed rotation frequency $\Omega$.  Note that $\Omega$ 
far exceeds the critical rotation frequency to support vortices 
$\Omega_{\rm c}$, such that the lowest energy state of the fluid 
is a vortex lattice.

The non-dimensionalisation of the GPE is based on the natural units 
of the homogeneous fluid \cite{Barenghi2016}:
the unit of length is the healing length $\xi$, the unit of speed
is $c$, the unit of time is 
$\tau=\xi/c=\hbar/\mu_0$, the unit of energy is $\mu_0$, 
and the unit of density is $n_0$.  Both our 3D and 2D numerical simulations 
are performed using XMDS2 \cite{xmds2}, an open-source partial and 
ordinary differential equation solver. 
The time evolution of the dimensionless GPE is computed
via an adaptive fourth-fifth order Runge-Kutta integration scheme 
with typical time step $dt=0.01 \tau$ and grid spacing $dx = 0.4 \xi$; 
these discretization numbers are sufficiently small to resolve 
the smallest spatial features (vortices and the fluid boundary layer, 
which are of the order of few healing lengths) and the shortest 
timescales in the fluid.  We typically conduct our 3D simulations 
on a cubic grid of size $256^3$.  Threaded parallel 
processing is employed using the OpenMP standard across typically $44$ 
threads to improve processing speeds on computationally intensive simulations.

\section{Results}
\label{section3}

\subsection{Typical spin-up dynamics} 

\begin{figure*}
(a) \hspace{3.8cm} (b) \hspace{4cm} (c) \hspace{5cm} \\
\includegraphics[width = 0.24\textwidth]{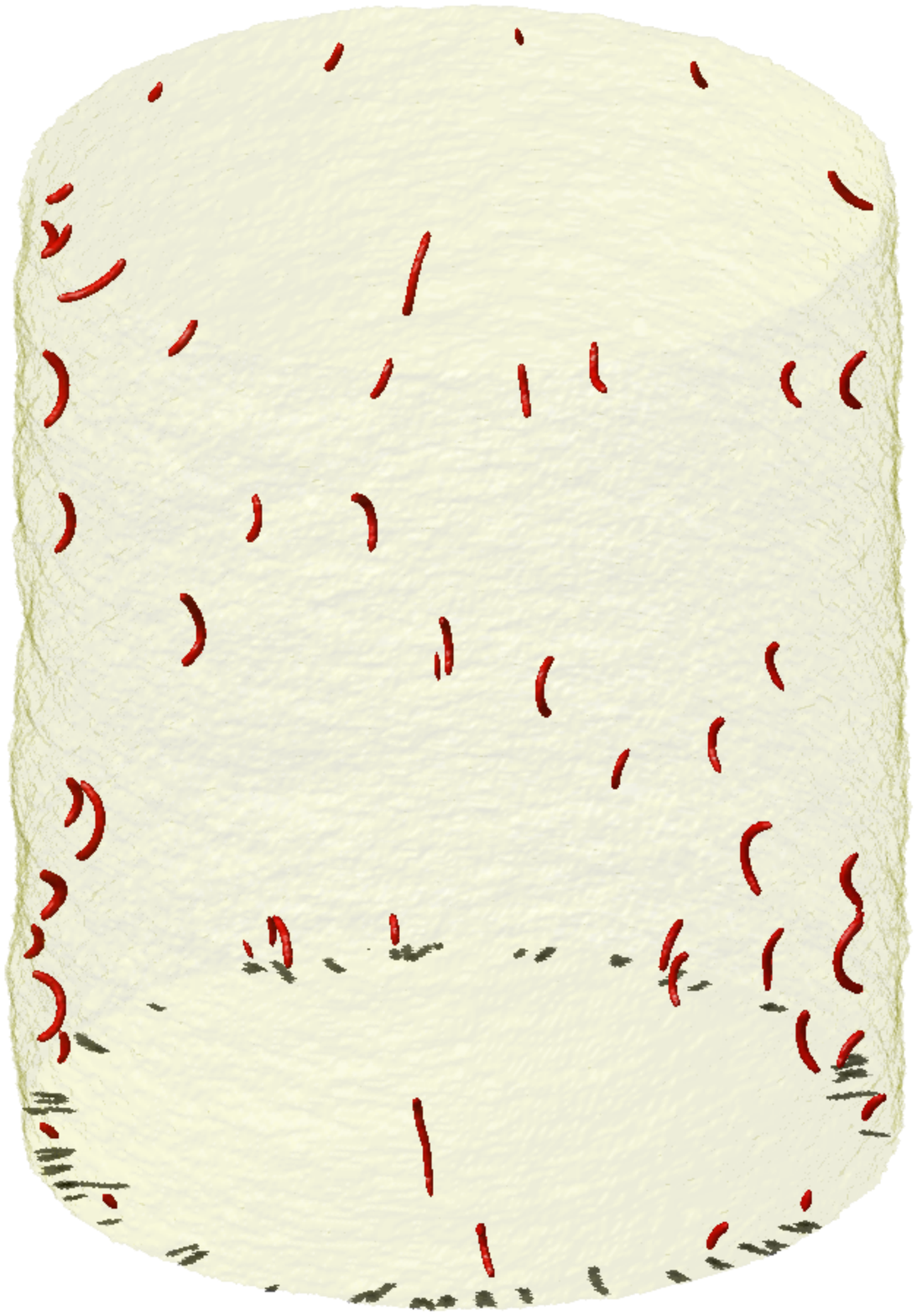}
\includegraphics[width = 0.24\textwidth]{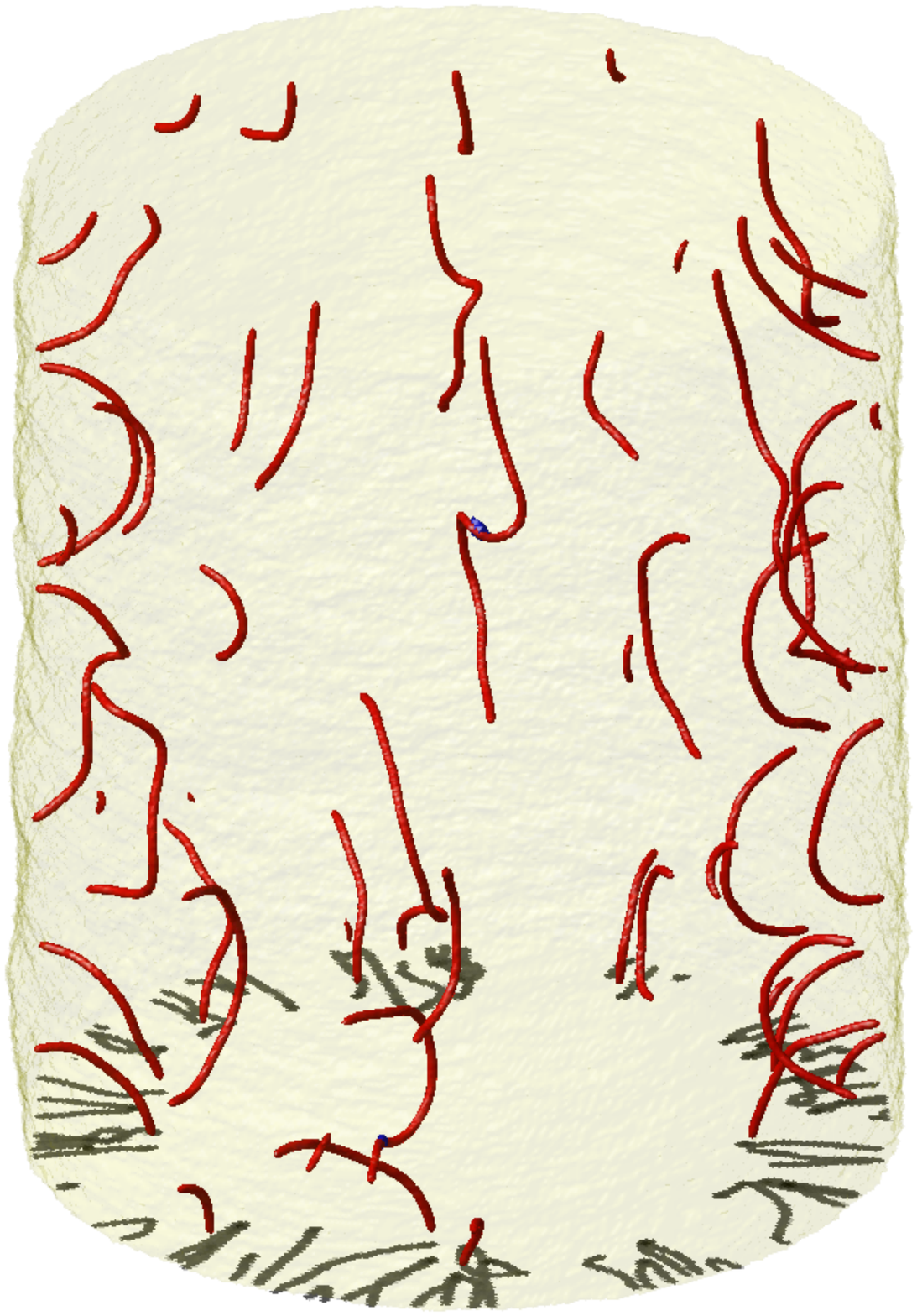}
\includegraphics[width = 0.24\textwidth]{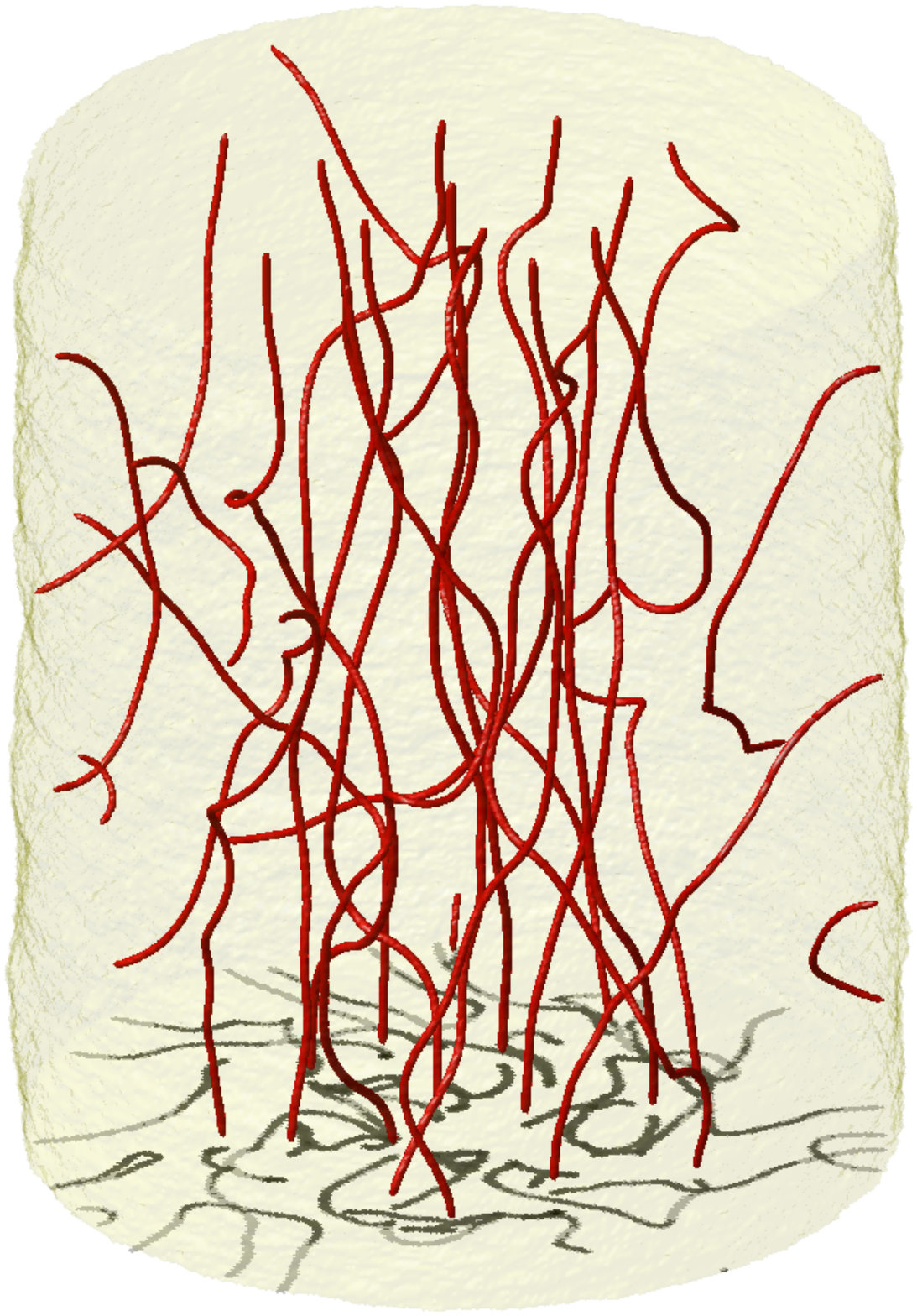}
\\(d) \hspace{4cm} (e) \hspace{3.8cm} (f) \hspace{3.8cm} \\
\includegraphics[width = 0.24\textwidth]{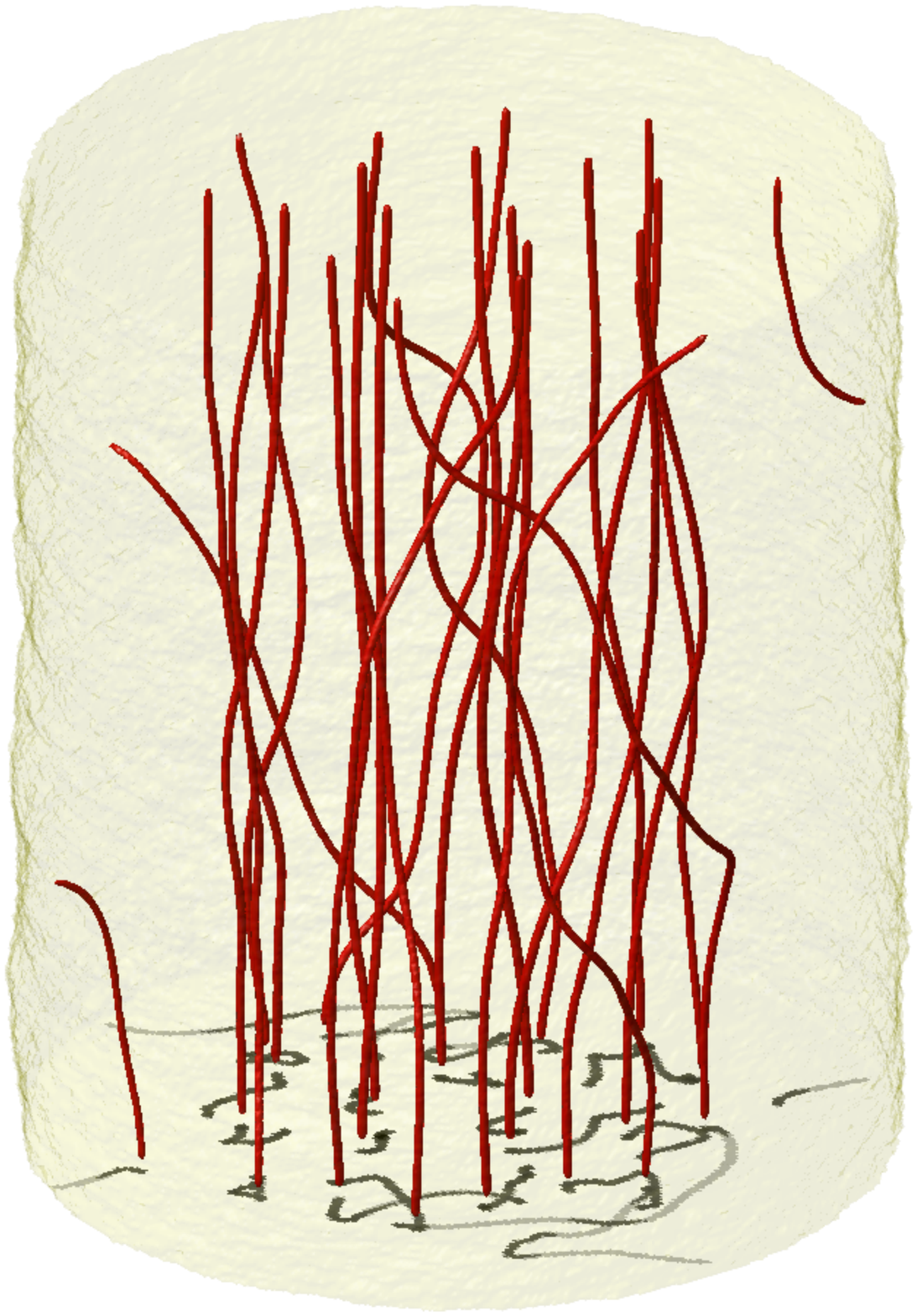}
\includegraphics[width = 0.24\textwidth]{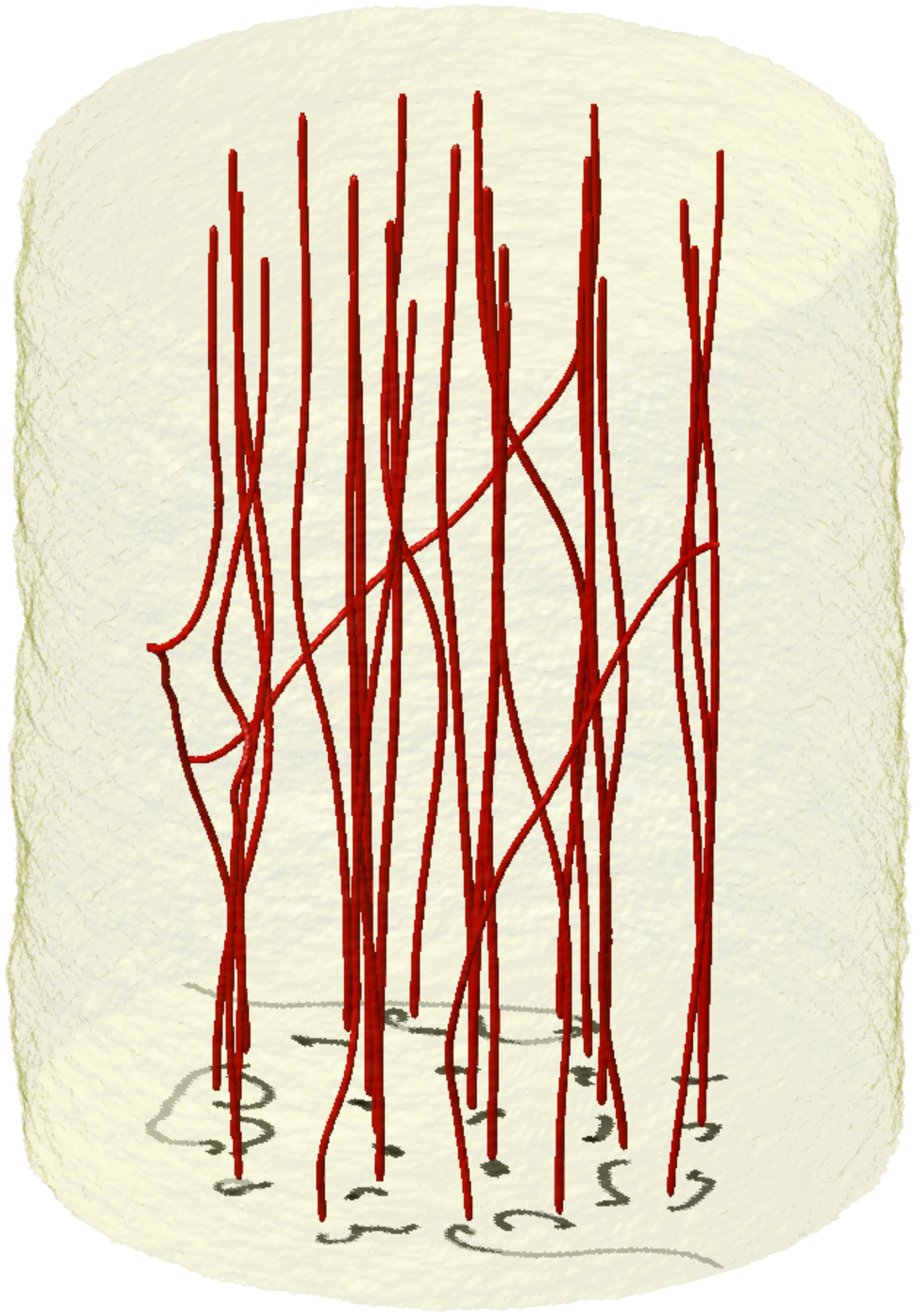}	
\includegraphics[width = 0.24\textwidth]{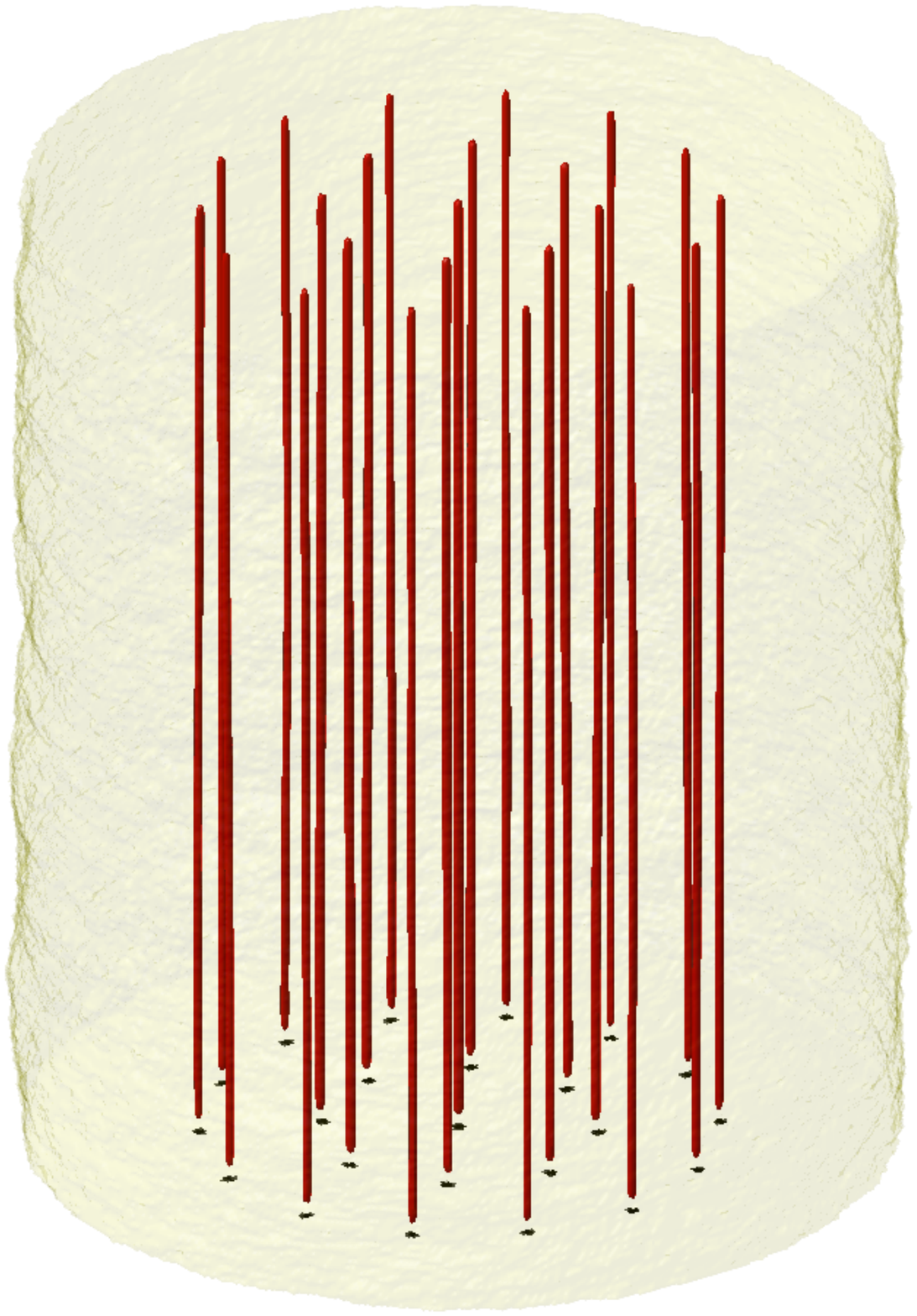}		
\caption{
Three-dimensional snapshots at times 
%$t/\tau=50$, $100$, $200$, $500$, $1000$ and $3000$ during the spin-up
$t/\tau=100$ (a), $200$ (b), $500$ (c), $1000$ (d), $1500$ (e) and $3000$ (f)
during the spin-up of the initially quiescent fluid. 
The vortex cores are identified by density isosurfaces;
vortices with positive and negative circulation (as determined by 
their pseudo-vorticity \cite{Villois2016} in the $z$ direction)
are visualised in red and blue respectively. The faint yellow
isosurface represents
the confining bucket. A false-color shadow is projected
onto the bottom surface to enhance the visualisation of the 3D vortex lines.
}
\label{fig2}
\end{figure*} 

We now demonstrate the typical spin up of an initially 
quiescent fluid.  Unless otherwise indicated, we present results
for the following choice of parameters: 
bucket radius $R=50\xi$, bucket height $H=100 \xi$, 
rotation frequency $\Omega = 0.02~ \tau^{-1}$,
dissipation parameter $\gamma=0.05$, and
roughness parameter $a=0.1$ (meaning that the irregular surface
of the bucket extends radially from $45\rm \xi$ to $50~\xi$, corresponding to
irregular `surface bumps' of height up to 5 healing lengths).

The evolution of the fluid is illustrated by the
snapshots shown in Fig.~\ref{fig2} in which the vortex lines are tracked
in 3D space using a precise method introduced in Ref. \cite{Villois2016}; movies of the evolution
are available in Supplementary Material \cite{movies}.  From the initial quiescent and vortex-free 
fluid, first we see the nucleation of vortex lines at the 
cylindrical boundary of the fluid [Fig.~\ref{fig2}(a)]. These vortices are all singly-quantized; we do not detect the presence of multiply-charged vortices in any of our simulations, which is consistent with the energetic instability of multiply-charged vortices and the favourability of singly-charged vortices \cite{Barenghi2016}.
The nucleation takes place at the sharpest features on the surface, as seen 
in a previous calculation over a flat rough surface \cite{Stagg2017}: at 
these features the local (potential) flow velocity is raised by the 
curvature of the boundary, and exceeds the critical velocity of 
vortex nucleation, which, according to Landau's criterion,
in a Bose gas is $v_c \approx c$.
Since the local flow speed around a moving obstacle always exceeds 
the translational speed of the obstacle, Landau's criterion can be
satisfied by a translational speed less than $c$.
For example, a cylindrical obstacle moving at speed approximately
equal to $0.4c$ will nucleate vortices \cite{Frisch1992,Stagg2014}. 
In our case ($\Omega =0.02/\tau$ and $R=50 \xi$), the translational 
speed of the prominences on the rough boundary
is approximately $\Omega R \approx c$, which is sufficient to exceed 
Landau's criterion and nucleate vortices.
Figure~\ref{fig2} (a) and (b) show that the
vortex lines which nucleate at the rough boundary have the shape of
small half-loops or handles; similar vortex shapes have been reported 
in trapped Bose-Einstein condensates
\cite{Aftalion2003} and turbulent superfluid helium-4 near
a heated cylinder \cite{Rickinson2020}, and have been called respectively
``U-vortices" and ``handles".  

\begin{figure}
\centering
\hspace{0.5cm}(a)\\
\includegraphics[width = 0.7\columnwidth]{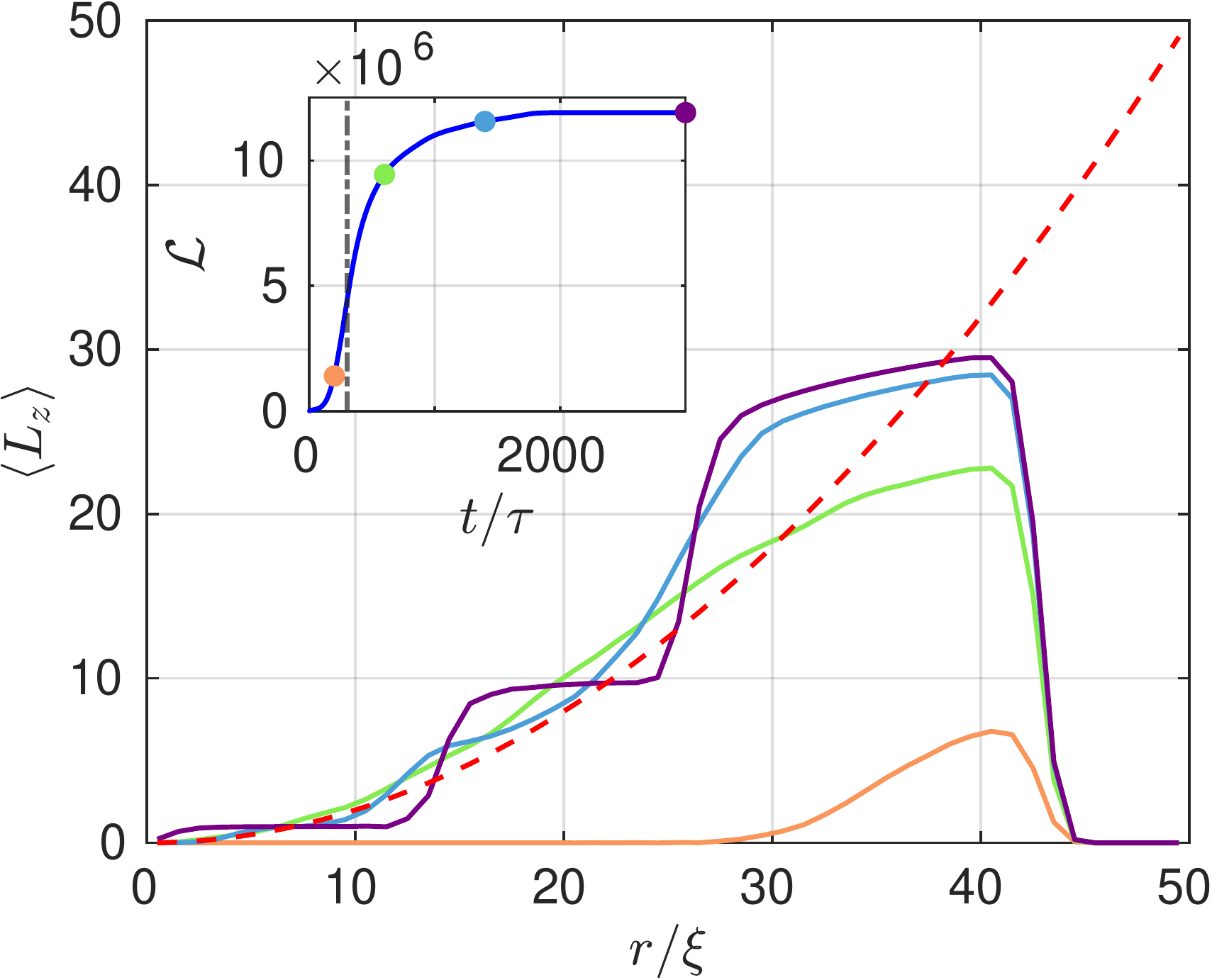}\\
\hspace{0.5cm}(b)\\
\includegraphics[width = 0.7\columnwidth]{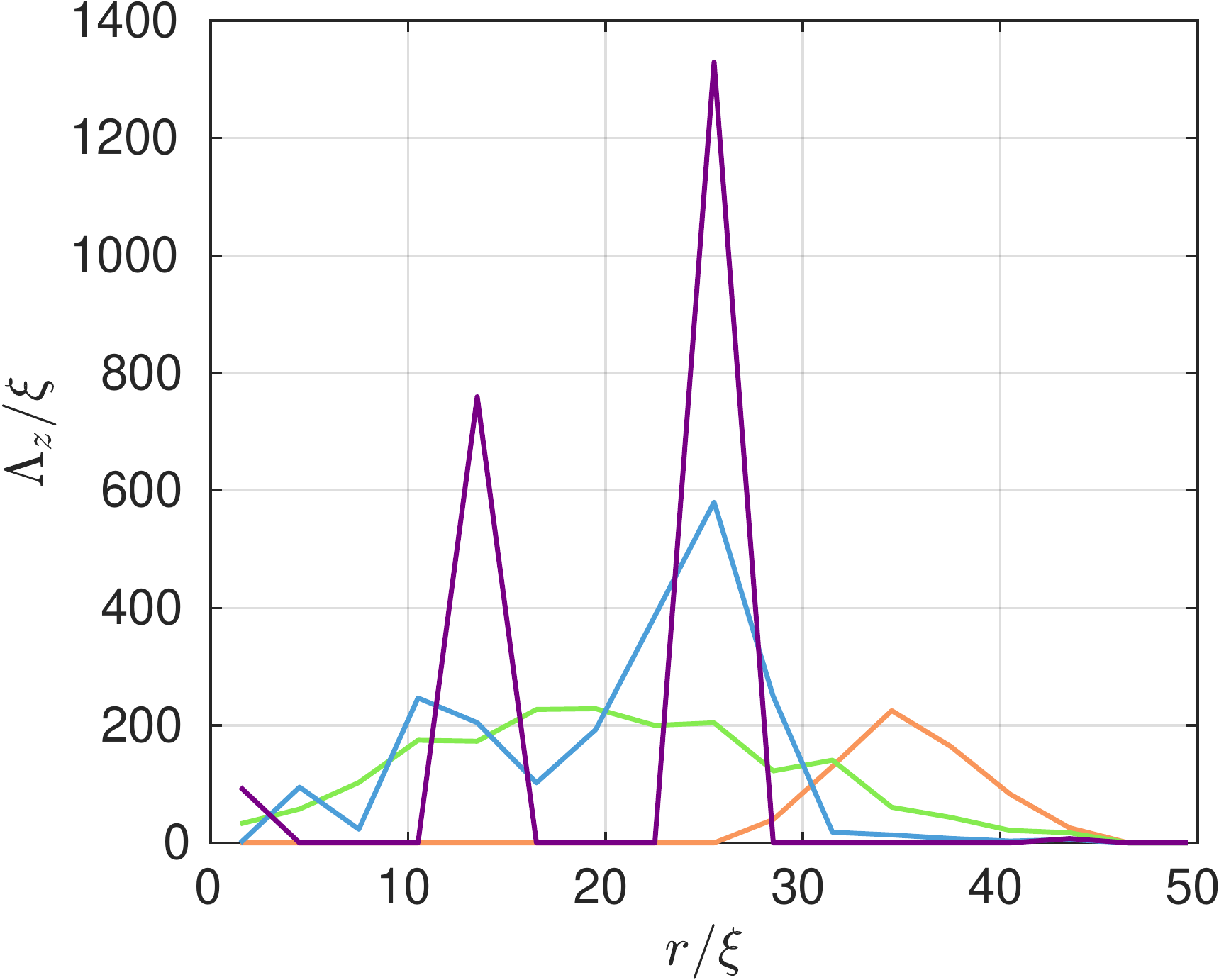}
\caption{(a) Angular momentum density as a function of radius at various times during the spin-up of the rough bucket (same parameters as in Fig. \ref{fig2}). Plotted is the angular momentum density averaged over the $z$-dimension, $\langle L_z \rangle$.    The lines correspond to times indicated by circular markers on the inset  The distribution of the angular momentum of a solid-body of uniform density is shown by the red dashed line.   The inset shows the evolution of the total angular momentum of the fluid $\mathcal{L}=\iiint L_z\, {\rm d}x\,{\rm d}y\,{\rm d}z$.  (b)  The vortex length projected in the $z$-direction 
$\Lambda_z$ plotted as a function of radius, at the same times as in (a).  The data is binned in radial intervals of $3\xi$.
}
\label{fig_Lz}
\end{figure}

We next consider the angular momentum of the fluid, exploring its evolution and distribution.  We define the density of the $z$-component of the angular momentum of the fluid as,
\begin{equation}
L_z(x,y,z) = \psi^* i\hbar\left(y\frac{\partial}{\partial x}-x\frac{\partial}{\partial y} \right) \psi.
\end{equation}
Figure \ref{fig_Lz}(a) shows the angular momentum density as a function of the radial coordinate, averaged vertically and azimuthally, $\langle L_z \rangle$.  At $t=0$, this is zero throughout the fluid.  At time evolves, angular momentum builds at the edge of the bucket and drifts inwards, corresponding to the nucleation and inward drift of the vortex lines.  At steady-state, the angular momentum forms a stepped curve, with each step corresponding to a concentric ring of vortex lines in the final lattice.  Note how the final distribution of the angular momentum density approximately follows the result of solid-body rotation with constant mass density, $m n_0 \Omega r^2$.  The total $z$-component of the angular momentum $\mathcal{L}=\iiint L_z\, {\rm d}x\,{\rm d}y\,{\rm d}z$  (inset in Fig. \ref{fig_Lz}(a)) grows in time, saturating at a final value by around $t\sim 2000~\tau$, which is when the vortex line have settled into the lattice configuration.  In gaseous superfluids confined within smooth potentials, recent results of merging superfluids \cite{Kanai2018,Kanai2020} suggest that the rate of angular momentum transfer between a static and rotating state is constant; however, here the growth of the angular momentum follows a sigmoidal curve, rather than a linear one.  %"the rate of the angular momentum transfer from the bucket wall to the BEC remains constant in the transient stage when the vortex tangle spreads out?"   

We also consider the distribution of the vortex length projected in the $z$-direction, 
$\Lambda_z$, as a function of radius, shown in Figure \ref{fig_Lz}(b).  At early times, vortex length exists only near the bucket edge, spreading progressively into the bulk.  Later, the vortex length converges towards falling at discrete peaks at $r=0$, $r\approx 13 \xi$ and $r\approx 26 \xi$, corresponding to the concentric arrangement of vortices in the lattice configuration.

\begin{figure}
\centering
\hspace{0.5cm}(a)\\
\includegraphics[width = 0.6\columnwidth]{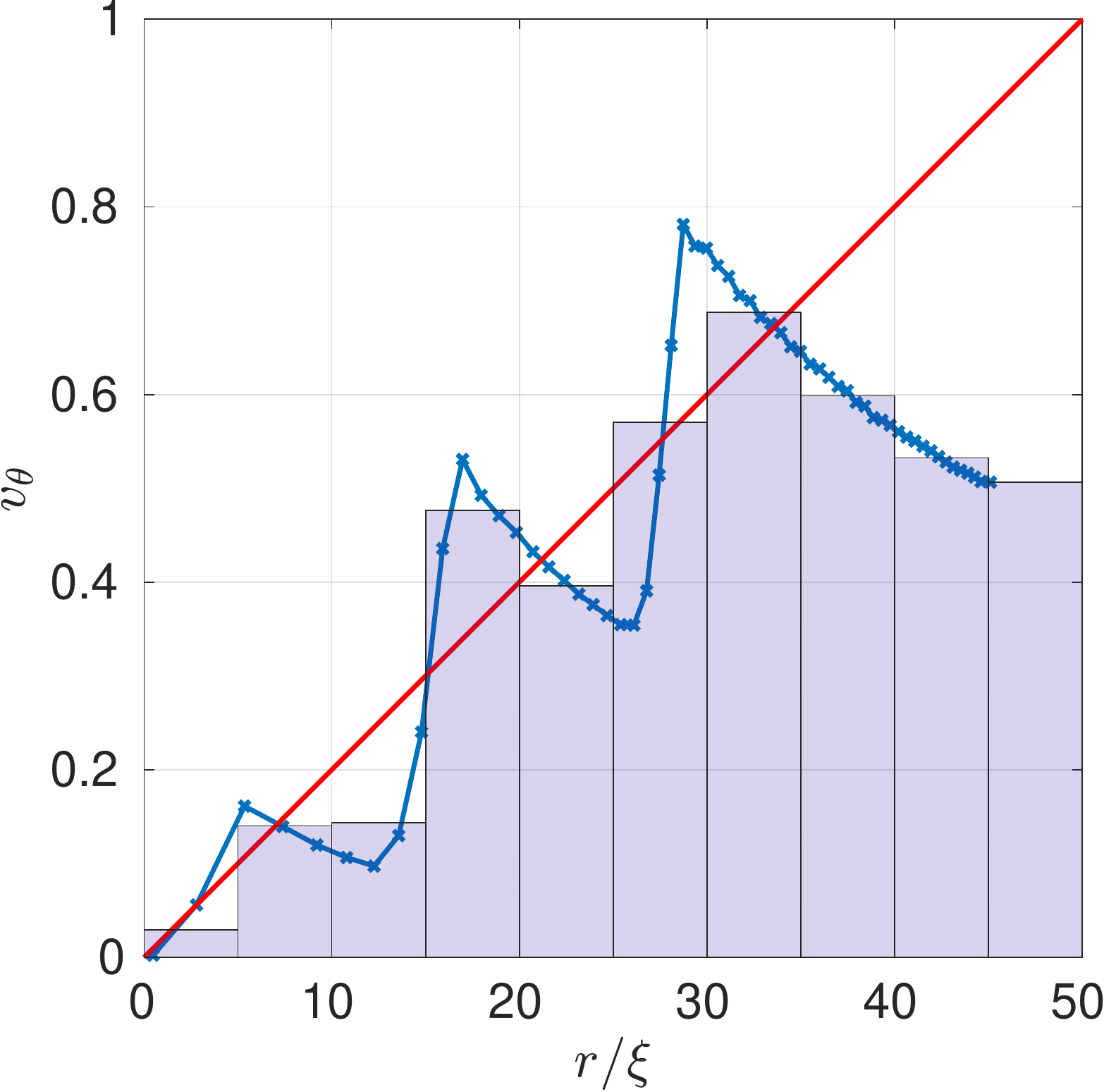}\\
\hspace{0.5cm}(b)\\
\includegraphics[width = 0.6\columnwidth]{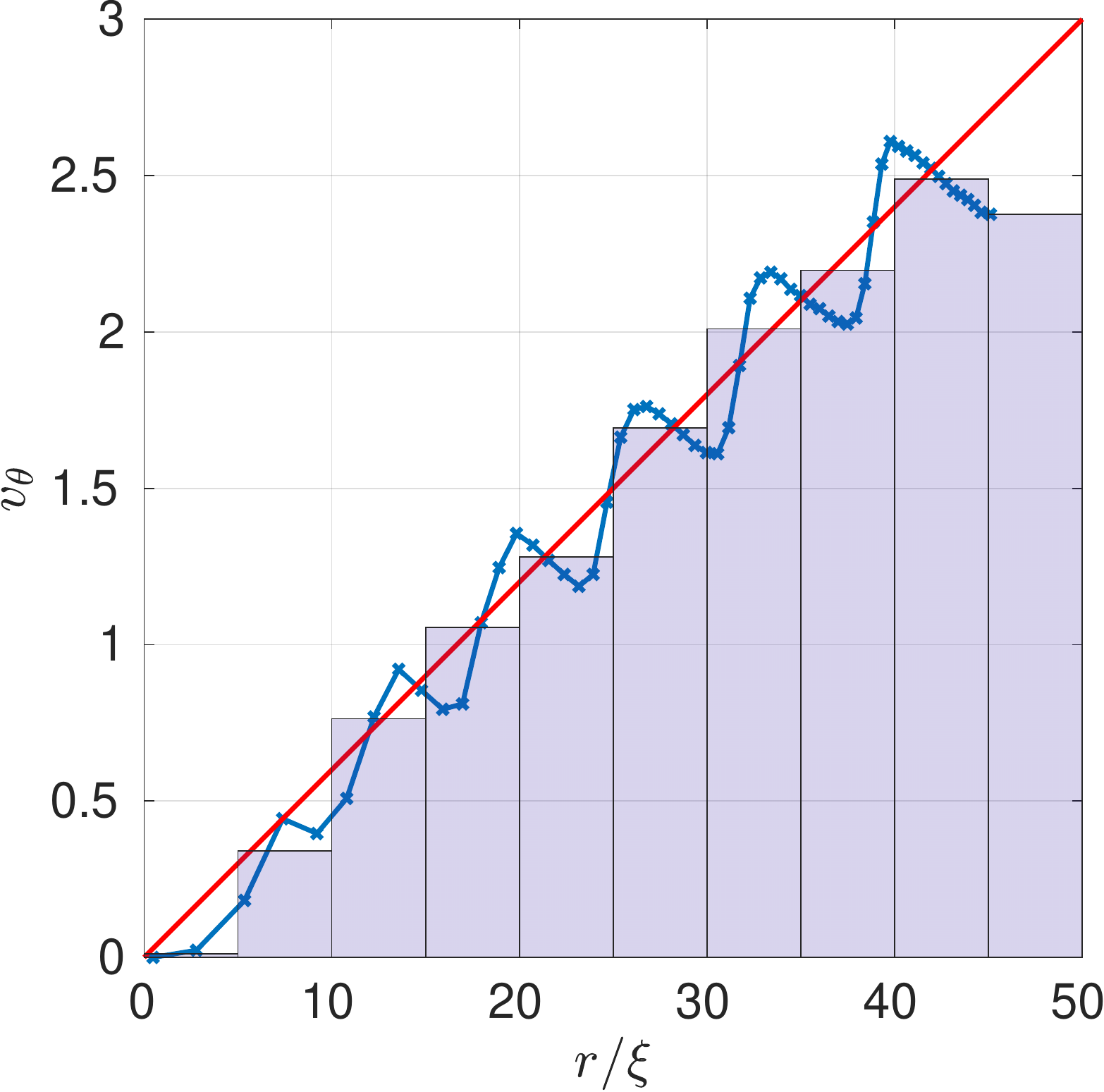}
\caption{Azimuthal velocity $v_{\theta}$ of the fluid as a function of radius $r$, for
rotation frequencies of $\Omega=0.02~\tau^{-1}$ (a) and
$\Omega=0.06~\tau^{-1}$ (b).  We use roughness parameter $a=0.1$.  The solid red lines 
represent solid body rotation $v_{\theta}=\Omega r$; the blue lines
are values of $v_{\theta}(r)$ averaged in the $\theta$ direction;
the pale blue rectangles are histograms with bin size $\Delta r=5 \xi$
(therefore the outer bins contain more data points). It is apparent that
the more rapid rotation (b) creates a vortex lattice in better agreement
with the solid body rotation, and that there is a vortex-free region
near the boundary.
}
\label{fig3}
\end{figure}

\begin{figure}
\centering
\includegraphics[width = 0.9\columnwidth]{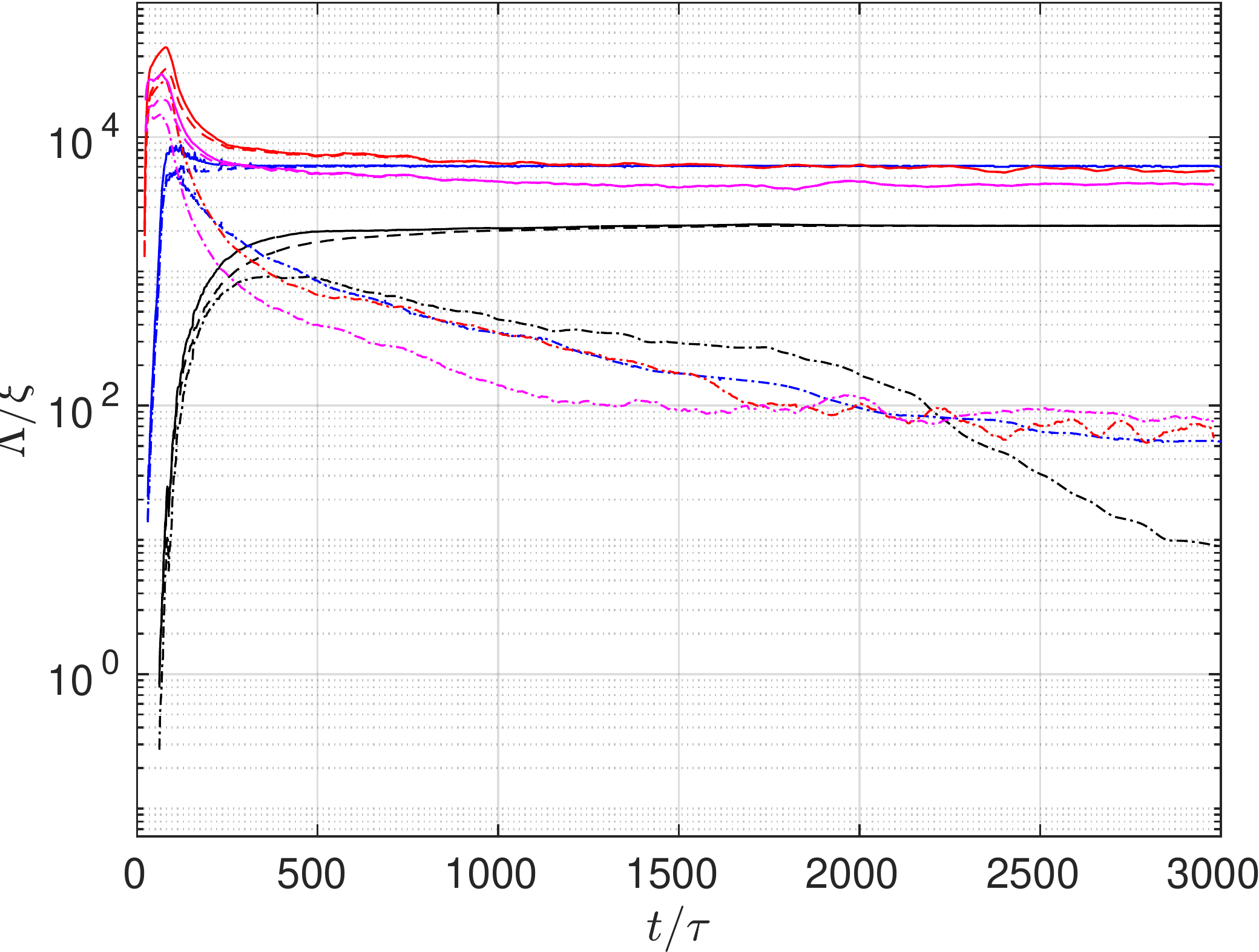}
\caption{
Evolution of the total vortex length, $\Lambda$ (solid lines), as well
as the vortex length in the $z$-direction, $\Lambda_z$ (dashed lines) and
the vortex length in the $xy$-plane, $\Lambda_{xy}$ (dot-dashed lines), 
plotted versus time $t$ for different angular velocity of rotation
$\Omega=0.02~\tau^{_1}$ (black), $0.04~\tau^{-1}$ (blue) and
$0.06 ~\tau^{-1}$ (red) 
%and $0.08~\rm \tau^{-1}$ (magenta), 
achieving final values of the vortex
length $\Lambda_{\infty}=2184 \xi$, $6007 \xi$ and $5568 \xi$ respectively.
%and $4701~\xi$ respectively.
All curves refer to roughness parameter $a=0.1$. 
}
\label{fig4}
\end{figure}

\begin{figure}
\centering
\includegraphics[width = 0.9\columnwidth]{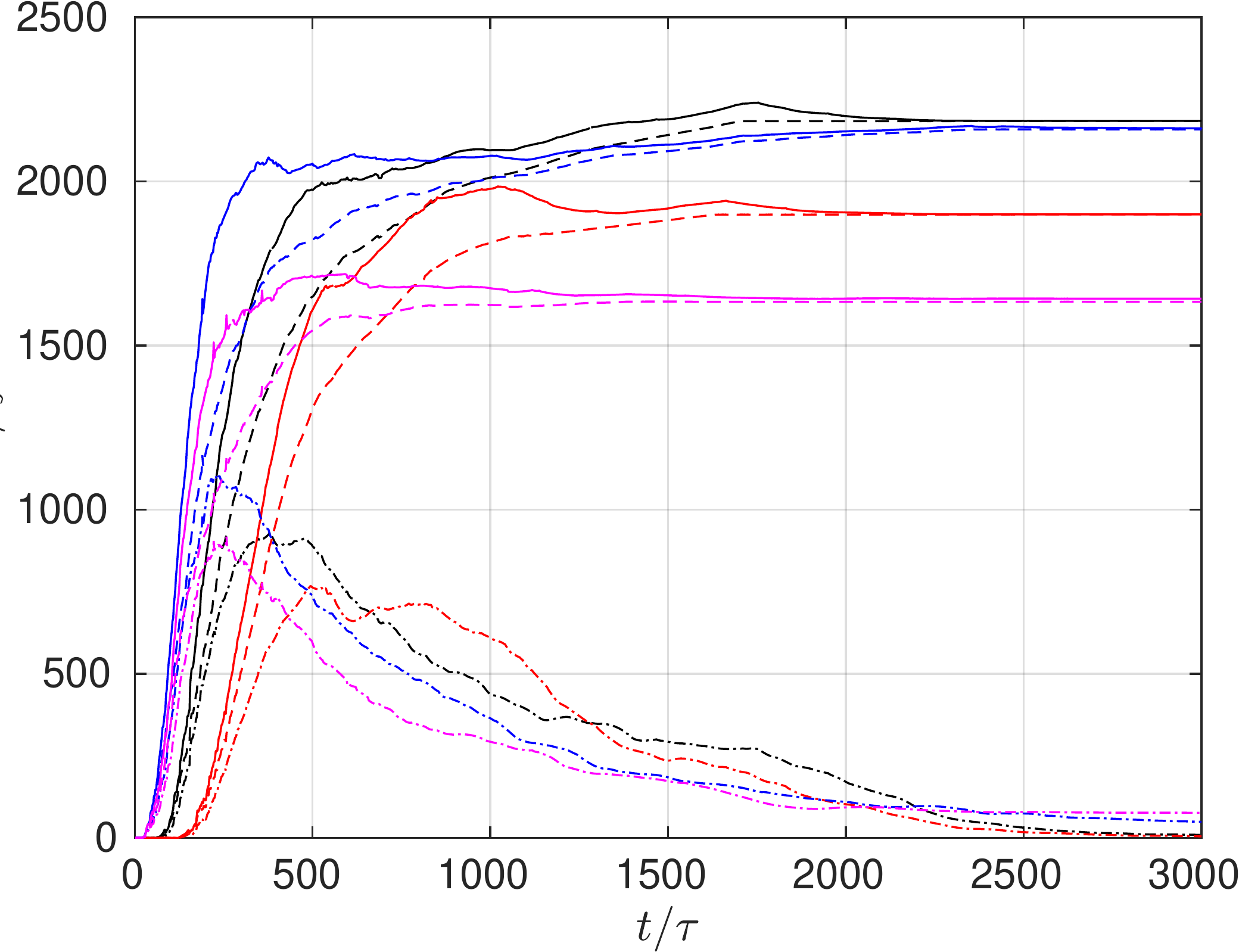}
\caption{
Evolution of the total vortex length, $\Lambda$ (solid lines), as well
as the vortex length in the $z$-direction, $\Lambda_z$ (dashed lines) and
the vortex length in the $xy$-plane, $\Lambda_{xy}$ (dot-dashed lines), 
plotted versus time $t$ for different boundary roughness
$a=0.05$ (red), $0.1$ (black), $0.2$ (blue), and $0.3~\rm \xi$ (magenta),
achieving final values $\Lambda_{\infty}=1899~\xi$, $2184~\xi$,
$2161~\xi$ and $1643~\xi$. All curves refer to the same
angular velocity $\Omega=0.02~\tau^{-1}$. 
}
\label{fig5}
\end{figure}

The collection of U-vortices nucleated at the boundary
is the superfluid's analog of a boundary layer, the region separating the
rotating boundary from the still quiescent bulk of the fluid.
The U-vortices tend to be aligned along the $z$-direction, creating
a superflow in the same direction of the rotating boundary.
The vortex nucleation is therefore short-lived, since the nucleated
U-vortices reduce the relative motion between the fluid and the
boundary, suppressing further nucleations.
In time, the U-vortices grow in size and extend further into the 
fluid [Fig. \ref{fig2}(b,c)], ultimately filling the bulk  
[Fig. \ref{fig2}(d)].  During this stage of the evolution,
the U-vortices also grow in vertical extent in the $z$-direction, 
occasionally connecting and merging with each other, thus increasing
their vertical extent.  When the length in the $z$-direction becomes of the order 
of the bucket's height $H$, one or both vortex endpoints start sliding along
the smooth top and/or bottom of the bucket.
Once most of the vortex lines are fully extended from the top to the bottom
of the bucket, they quickly drift into the 
bulk of the fluid.  Although the vortex lines are aligned along the direction of
rotation, they remain highly excited and undergo reconnection events when
they collide with each other.  Over time they relax towards a regular 
configuration of straight vortices.  A small proportion of U-vortices
remain attached to the side of the bucket for a longer period of time 
[Fig.~\ref{fig2}(d)]; over a longer time they detach,
and relax to the final lattice configuration.  Some of the vortex lines
end up diagonally across the rest of the vortex lattice
[Fig.~\ref{fig2}(e)]: eventually they also relax to the final
lattice configuration  
[Fig. \ref{fig2}(f)].  The vortex lattice is stationary in 
the rotating frame, representing the lowest energy state of 
the rotating superfluid.  In this final state, the coarse-grained 
fluid velocity approximates the solid-body result 
${\bf v} = v_{\theta}\, {\bf e_\theta}= \Omega \, r \, {\bf e_\theta}$, 
where ${\bf e_\phi}$ is the azimuthal unit vector, as shown in Fig.~\ref{fig3};
as expected, the agreement improves with increasing $\Omega$, and there is
a vortex-free region near the boundary. 

Our 3D results are presented for a fixed bucket size due to computational constraints of simulating a larger system.  For a larger bucket we would expect qualitatively similar dynamics; indeed our 2D results in a larger bucket presented in Section IV C support this.  The most significant change under a larger bucket is more vortices in the final state (at a fixed rotation frequency) and as a result a better approximation to solid-body rotation.

\subsection{Role of angular velocity and roughness} 

To analyse the vortex dynamics further it is useful to distinguish the total vortex length, 
$\Lambda$, from the vortex length projected in the $z$-direction, 
$\Lambda_z$, and the vortex length projected in the $xy$-plane, 
$\Lambda_{xy}$. In the final vortex lattice all vortex lines 
are aligned along $z$, hence we expect that, after a sufficiently long time, $\Lambda_{xy}\approx 0$ and
$\Lambda_z \approx \Lambda$, with $\Lambda \rightarrow N_v H$, where $N_v$ is the final number of straight vortex lines.  
Figure~\ref{fig4} displays $\Lambda$ (solid lines), $\Lambda_z$ (dashed lines)
and $\Lambda_{xy}$ (dot-dashed lines) as a function of time for different
angular velocities of rotation, $\Omega=0.02$, $0.04$ and $0.06$
% and $0.08~\tau^{-1}$ 
at the same roughness parameter $a=0.1$. 
 It is apparent that in the initial stage, 
a great amount of vorticity is in the $xy$-plane, before realignment 
of the vortex lines along the $z$-axis of rotation takes place. The effect
is particularly noticeable at the largest angular velocities, for which,
during the initial transient, the vortex length is considerably larger
than the value $\Lambda_{\infty}$ achieved in the final vortex lattice
configuration. Moreover, we see that the final vortex line length increases with $\Omega$ due to the increasing number of vortices in the final state. 

Figure~\ref{fig5} shows $\Lambda$, $\Lambda_z$ and $\Lambda_{xy}$ 
plotted versus time at the same angular velocity $\Omega=0.02~\tau^{-1}$ 
for different values of 
roughness parameter $a$. The largest values of the final vortex
length $\Lambda_{\infty}$ are achieved with $a=0.1 \xi$ and $a=0.2\xi$. 
Smoother ($a=0.05\xi$) and rougher ($a=0.3\xi$) boundaries generate 
less vortex length. These variations in the final line length arise to the final number of vortex lines varying by a few vortices across these cases.
It is not surprising that the final vortex lattice depends
on the roughness which has nucleated the initial 
vorticity. Feynman's rule [Eq.~\eqref{eq:Feynman}] only refers to 
an idealised homogeneous system. Boundaries are known to have effects 
(e.g. missing vortex lines near the boundary) and 
it has been observed that the formation of the vortex lattice may be 
history-dependent and involve
metastability \cite{CampbellZiff,Wood2019} and hysteresis \cite{Mathieu}. 

\begin{figure}
\centering
\includegraphics[width = 0.8\columnwidth]{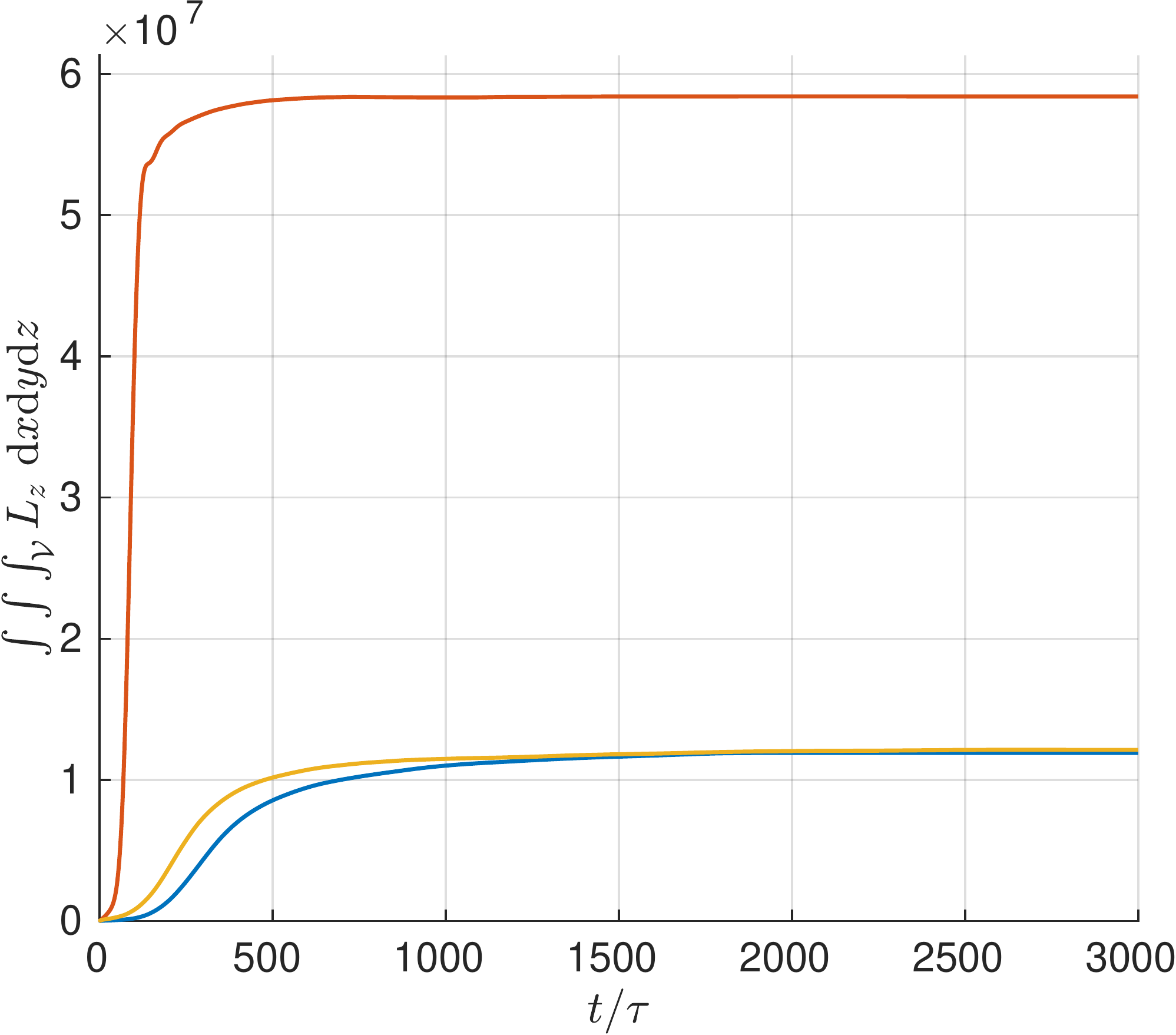}\\
\caption{
Evolution of the total angular momentum for the default parameters used in Fig. \ref{fig2} (blue line), double the rotation frequency (red line) and double the roughness amplitude (yellow line).  
}
\label{fig_Lz_total}
\end{figure}

Figure~\ref{fig_Lz_total} compares the growth of angular momentum between the default case (blue line), the case where the rotation frequency is doubled (red line), and the case where the roughness amplitude is doubled (yellow line).  The growth behaviour is qualitatively similar in all cases.  Doubling the rotation frequency leads to a much faster rate of injection of angular momentum, and a higher final value, consistent with the faster injection rate of vortex lines from the boundary and the higher density of vortex lines in the final lattice state.  Doubling the surface roughness has little effect on the growth of the angular momentum, just slightly increasing the rate of angular momentum injection, which can be attributed to the greater injection rate of vortices from the rougher surface.

\begin{figure*}
(a) \hspace{3.8cm} (b) \hspace{4cm} (c) \hspace{7cm} \\
	\includegraphics[width = 0.24\textwidth]{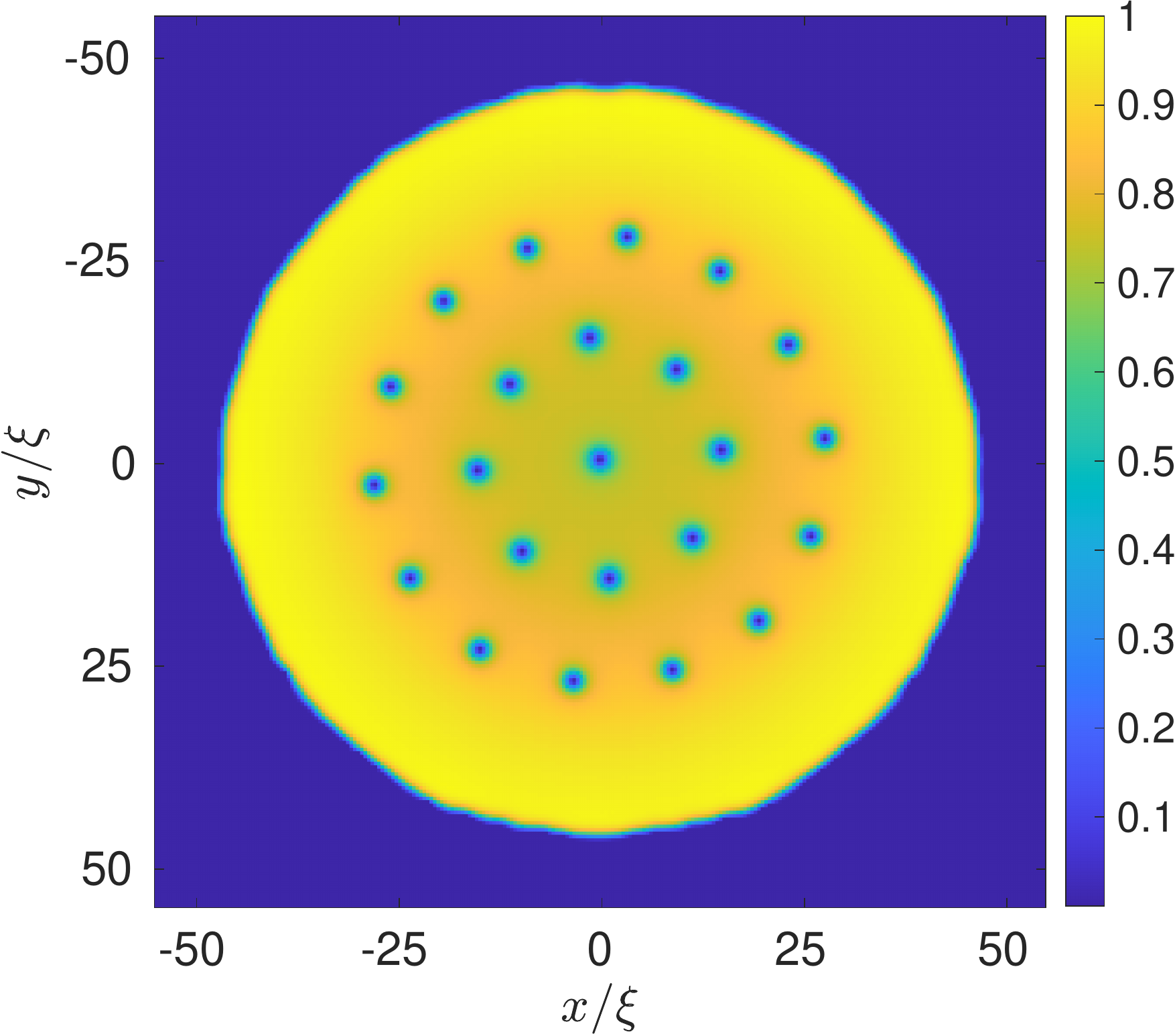}
        \includegraphics[width = 0.24\textwidth]{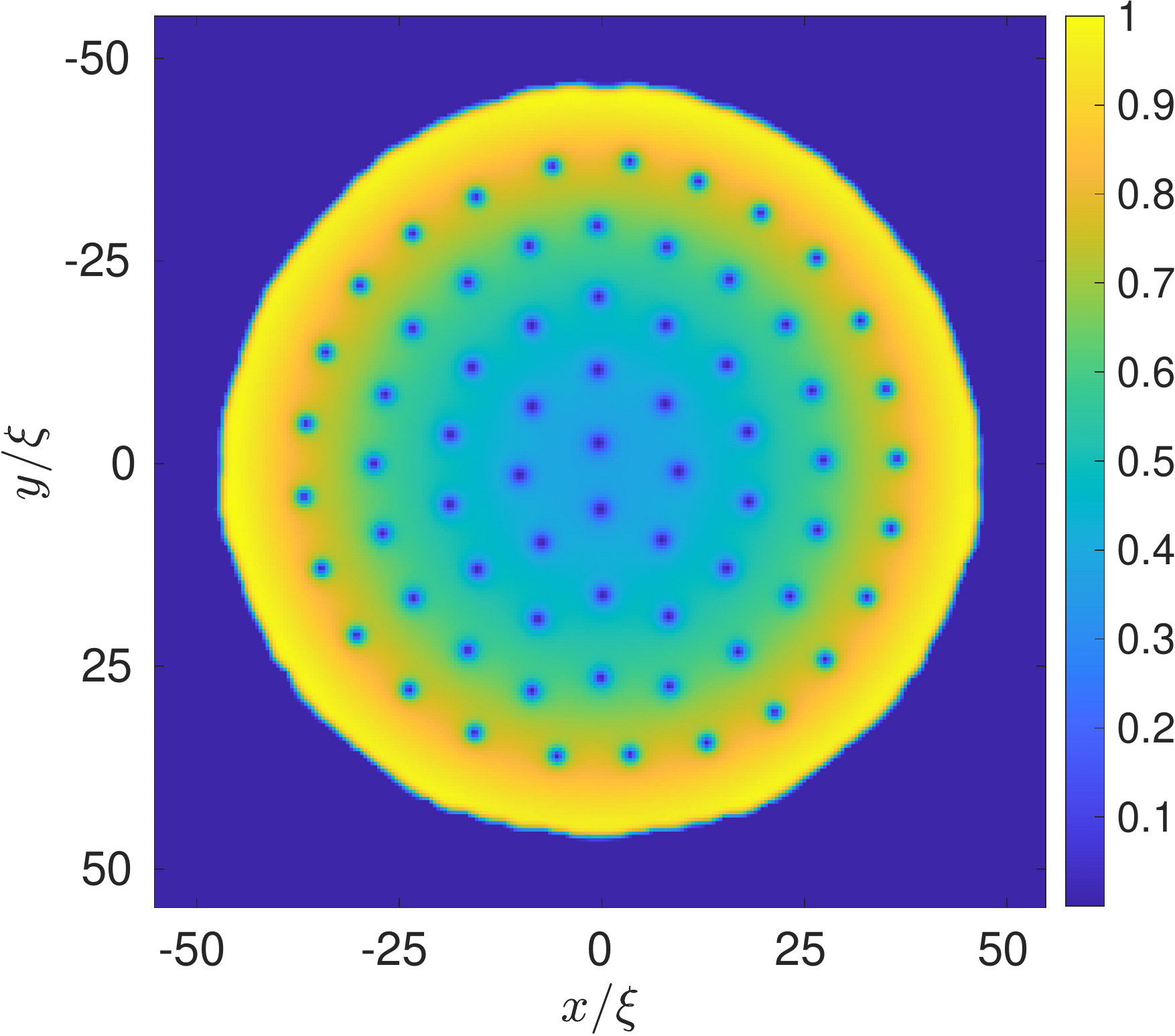}
	\includegraphics[width = 0.24\textwidth]{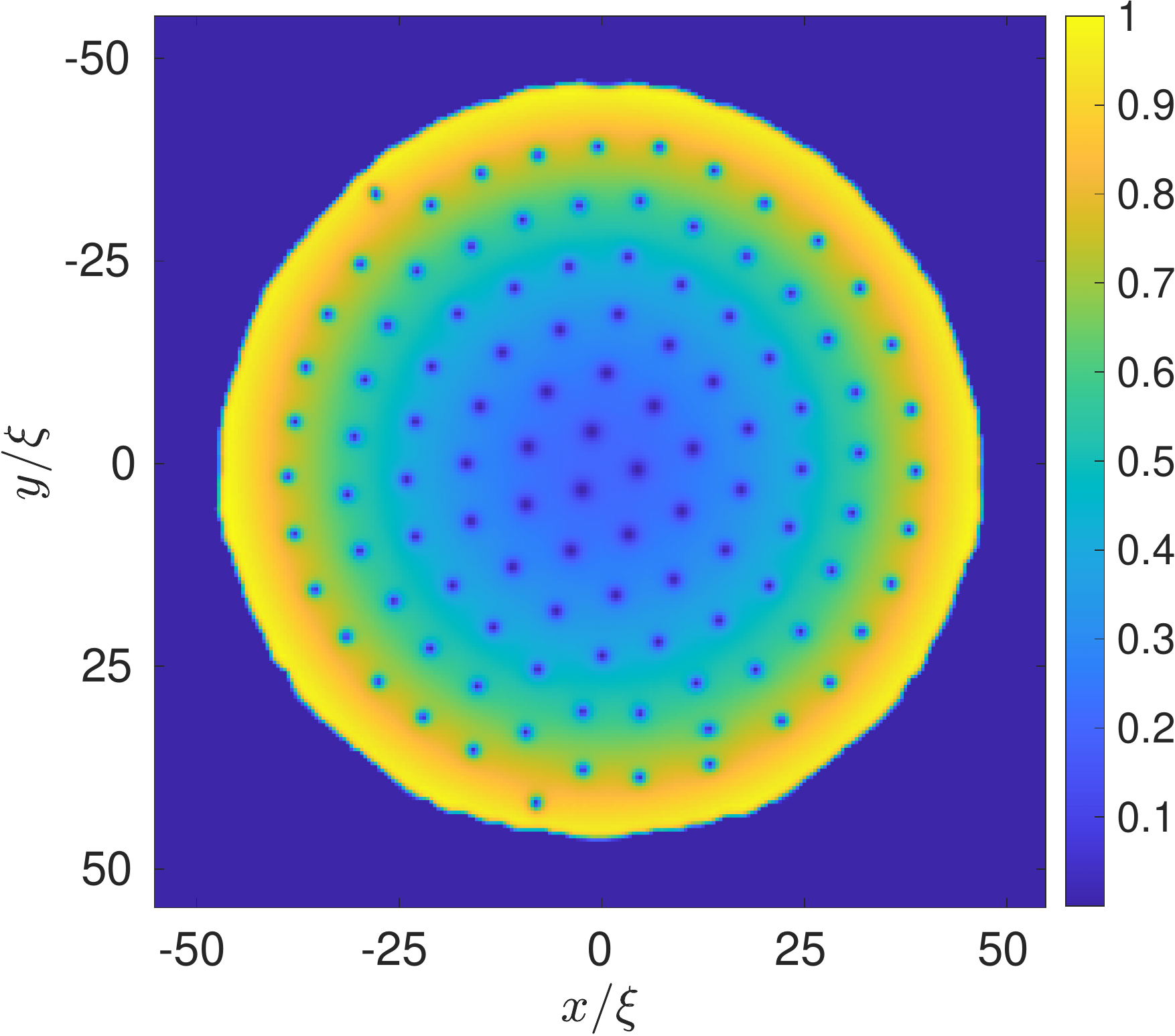}
\\(d) \hspace{4cm} (e) \hspace{3.8cm} (f) \hspace{3.8cm} \\
	\includegraphics[width = 0.24\textwidth]{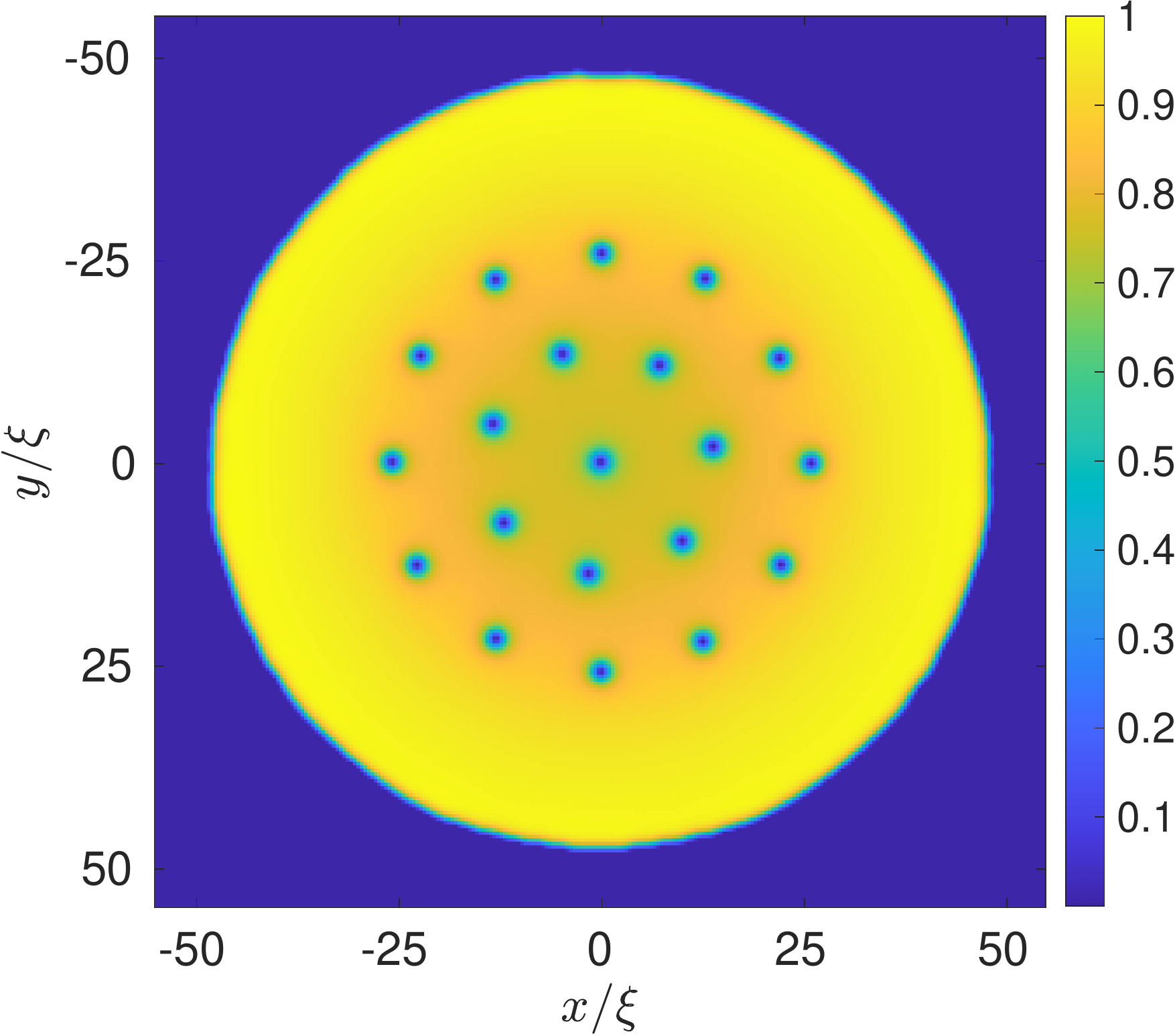}
	\includegraphics[width = 0.24\textwidth]{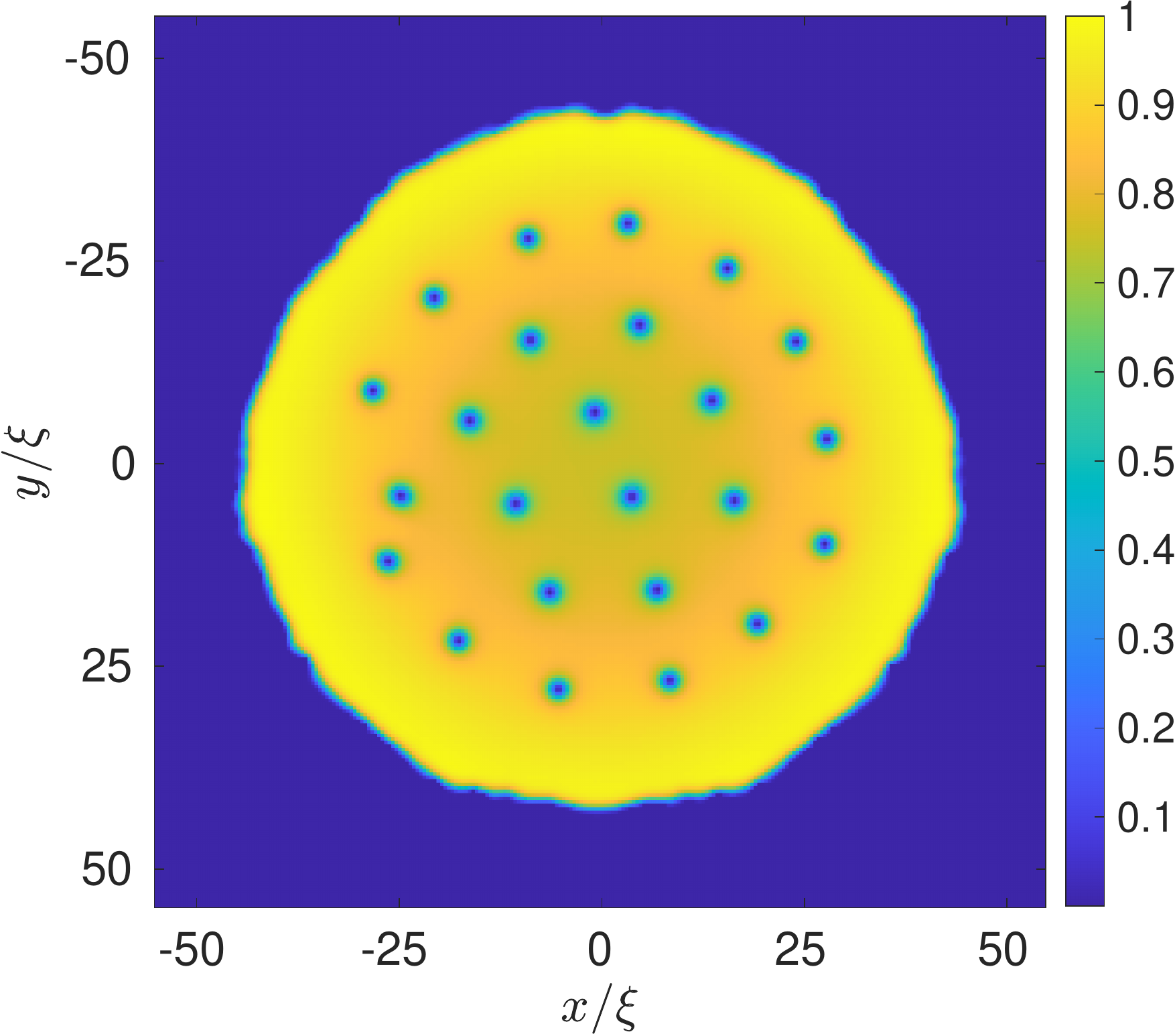}
	\includegraphics[width = 0.24\textwidth]{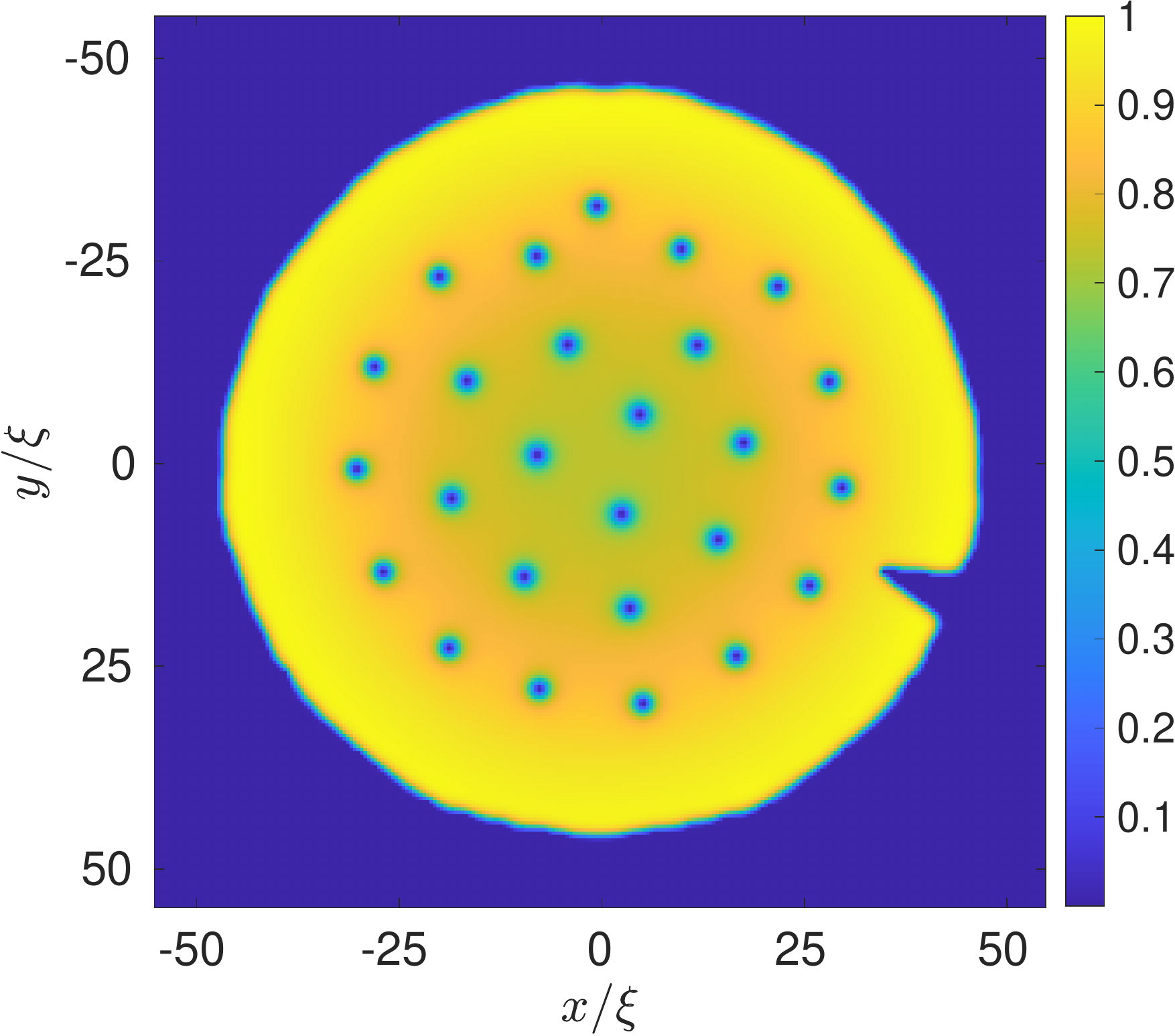}	
\caption{
Density profile in the $xy$-plane at the half-height of the bucket, showing the final vortex configurations for the following cases:
(a): $\Omega=0.02~\tau^{-1}$ and $a=0.1$ (default choice);
(b): $\Omega=0.04~\tau^{-1}$ and $a=0.1$ (double rotation);
(c): $\Omega=0.06~\tau^{-1}$ and $a=0.1$ (triple rotation);
(d): $\Omega=0.02~\tau^{-1}$ and $a=0.05$ (half roughness);
(e): $\Omega=0.02~\tau^{-1}$ and $a=0.2$ (double roughness);
(f): $\Omega=0.02~\tau^{-1}$ and $a=0.1$ (single strong protuberance added,
see Section~\ref{section4}).  
}
\label{fig6}
\end{figure*}

Figure~\ref{fig6} illustrates some of the final vortex patterns which we have
computed by plotting the superfluid density,
$\vert \psi(x,y) \vert^2$ in the $xy$-plane at
half-height of the bucket.  In these pictures
the vortices appear as small holes; to clarify the lengthscales,
we recall that on the vortex axis the density is zero and that
at distance $r=2 \xi$ from the axis, the density recovers about $80 \%$ of the
bulk value at infinity.
%It must be stressed that the figure shows slices of 3D vortex configurations, 
%not vortex patterns. Note that at half-height the vortex lines are as
%far as possible from the top and bottom boundaries, therefore more likely
%to be bent away from the ideal 2D vortex pattern. 
It is interesting
to compare the different final vortex configurations for halved/doubled rotation velocity and the roughness parameter with respect
to our default choice ($\Omega=0.02~\tau^{-1}$ and $a=0.1$).  While the ideal
2D vortex lattice has a vortex at the centre, surrounded by a first row of 6 vortices,
a second row of 12 vortices, etc, the vortex configurations shown
in Fig.~\ref{fig6} contain slightly different vortex numbers; in
particular some configurations contain vortex lines which seem 
misplaced [Fig.~\ref{fig6}(c)]
or lack the vortex at the centre [Fig.~\ref{fig6}(e)];
these configurations are metastable states corresponding to local 
minima of the free energy in the rotating frame \cite{CampbellZiff}. 
Moreover, at slow rotations [Fig.~\ref{fig6}(a,d)] the predicted vortex-free region near the boundary 
\cite{NorthbyDonnelly1970,ShenkMehl1971,StaufferFetter1968} is clearly
visible; this phenomenon affects the coarse-grained azimuthal velocity near the
boundary shown previously in Fig.~\ref{fig3}(a). The depletion of the background fluid density in the centre of the bucket - particularly evident in Fig. \ref{fig6}(b) and (c) - is due to coarse-grained centrifugal effects, analogous to the classical rotating case \cite{Barenghi2016}.

\section{Other effects}
\label{section4}

In this section we
repeat the simulation of Section~\ref{section3} with several 
significantly modifications: the presence of a single strong
protuberance, the presence of remanent vortex lines, and the 2D case.
The aim is to determine whether these effects change qualitatively the dynamics 
described in Section~\ref{section3}.

\subsection{Effect of a strong protubance}

\begin{figure*}
\hspace{-0.2cm}  (a) \hspace{2.9cm} (b) \hspace{2.9cm}(c) \hspace{2.9cm} (d) \hspace{2.9cm} (e) \hspace{1cm}\\
		\includegraphics[width = 0.19\textwidth]{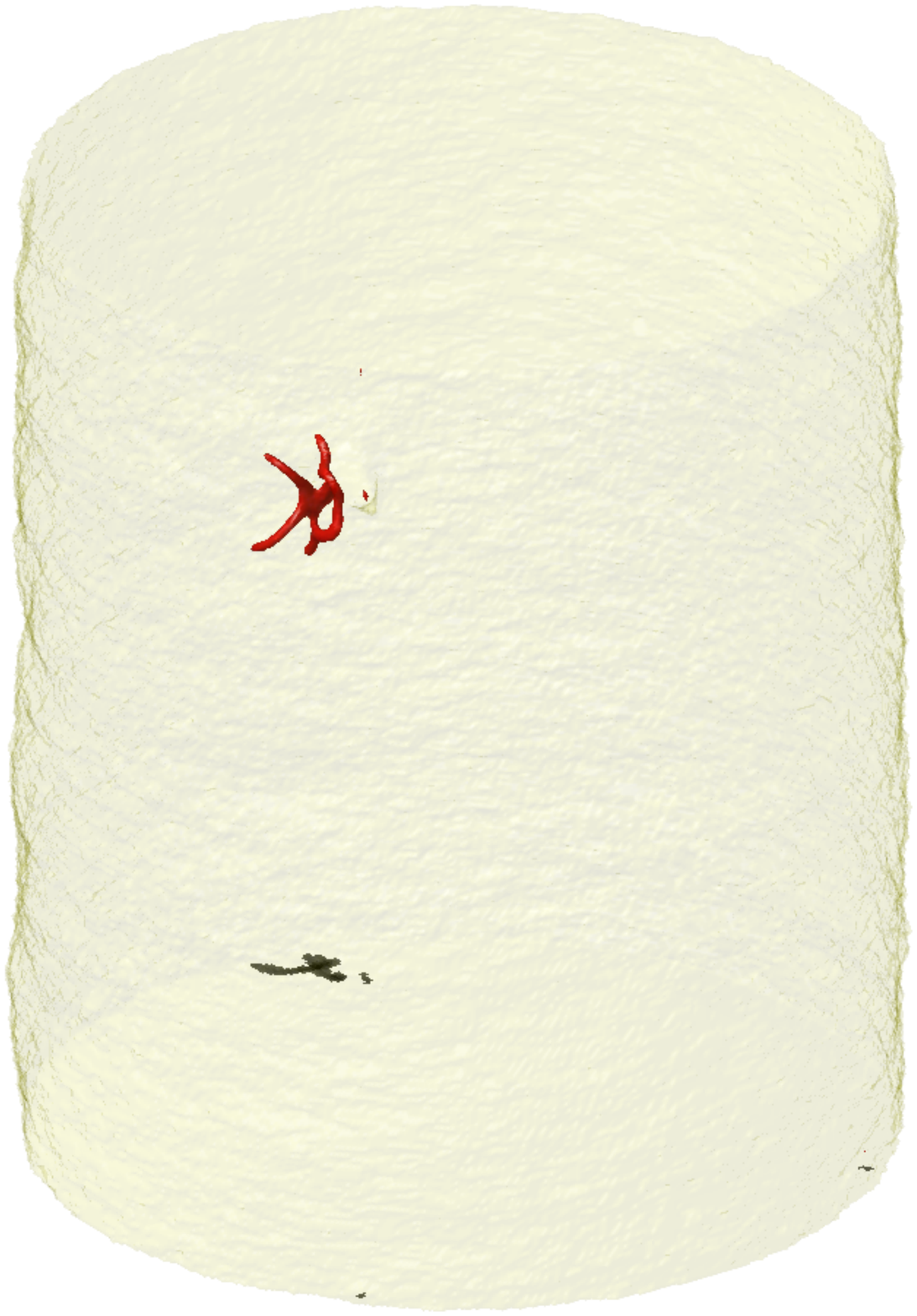}
		\includegraphics[width = 0.19\textwidth]{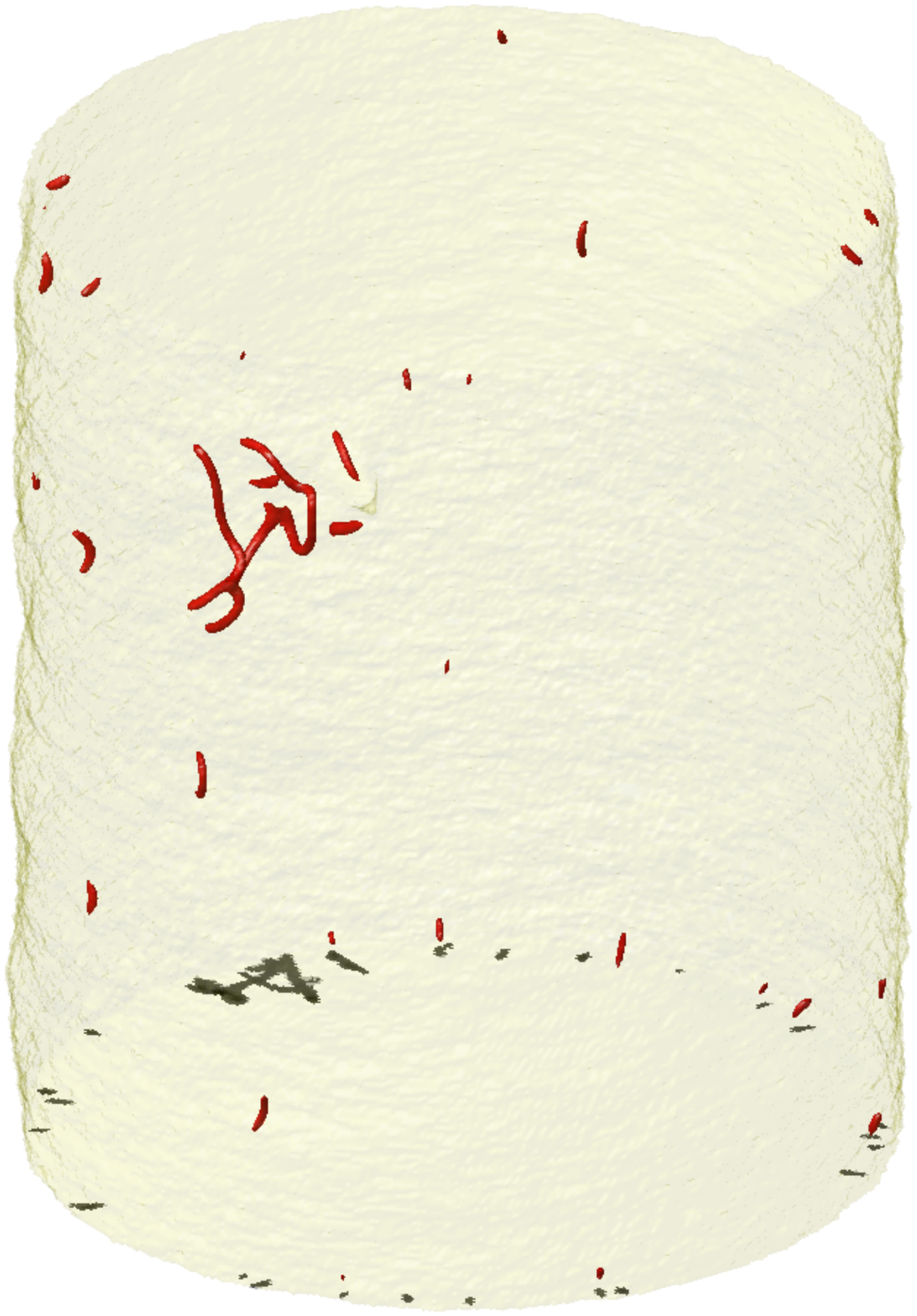}
		\includegraphics[width = 0.19\textwidth]{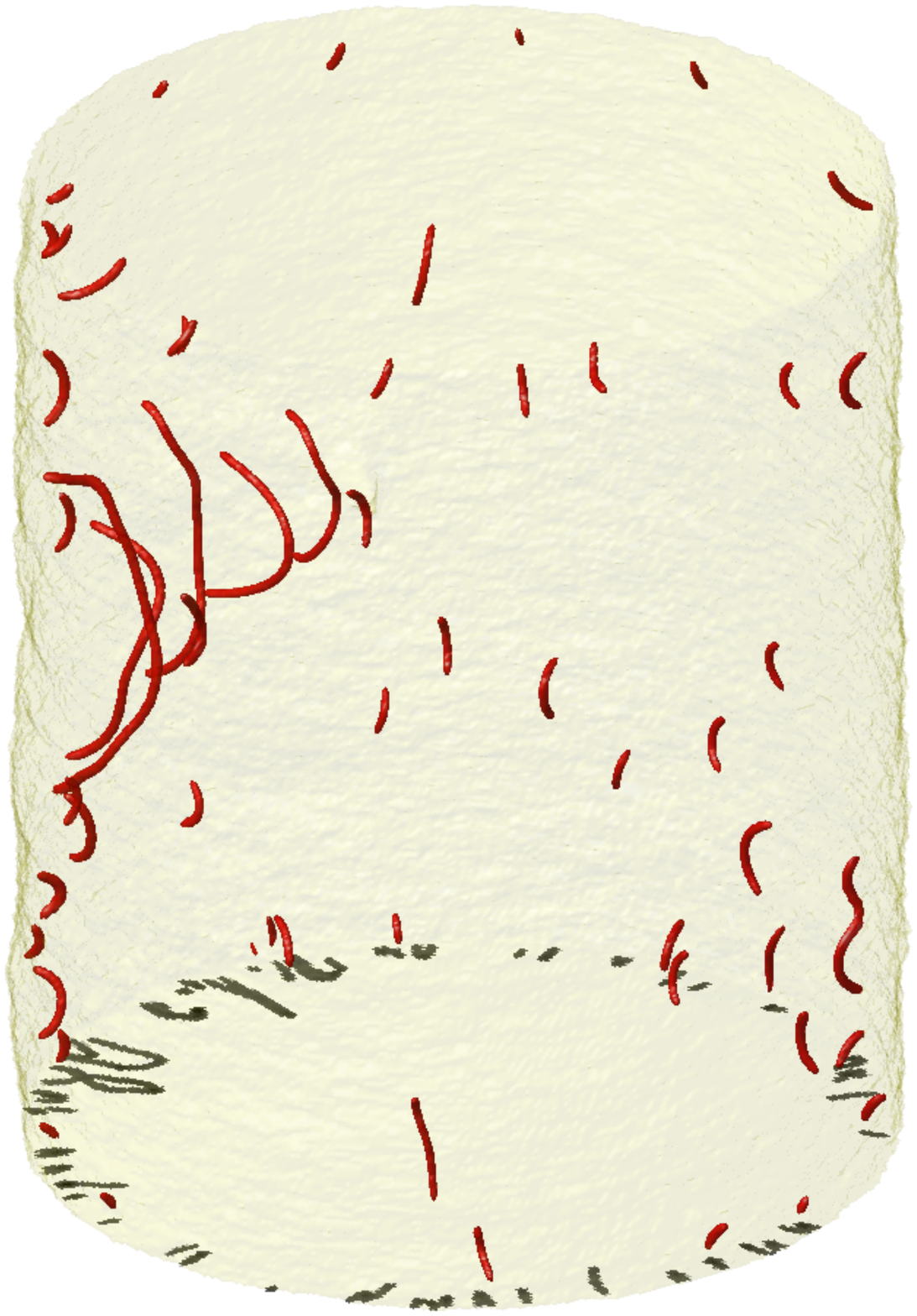}
		\includegraphics[width = 0.19\textwidth]{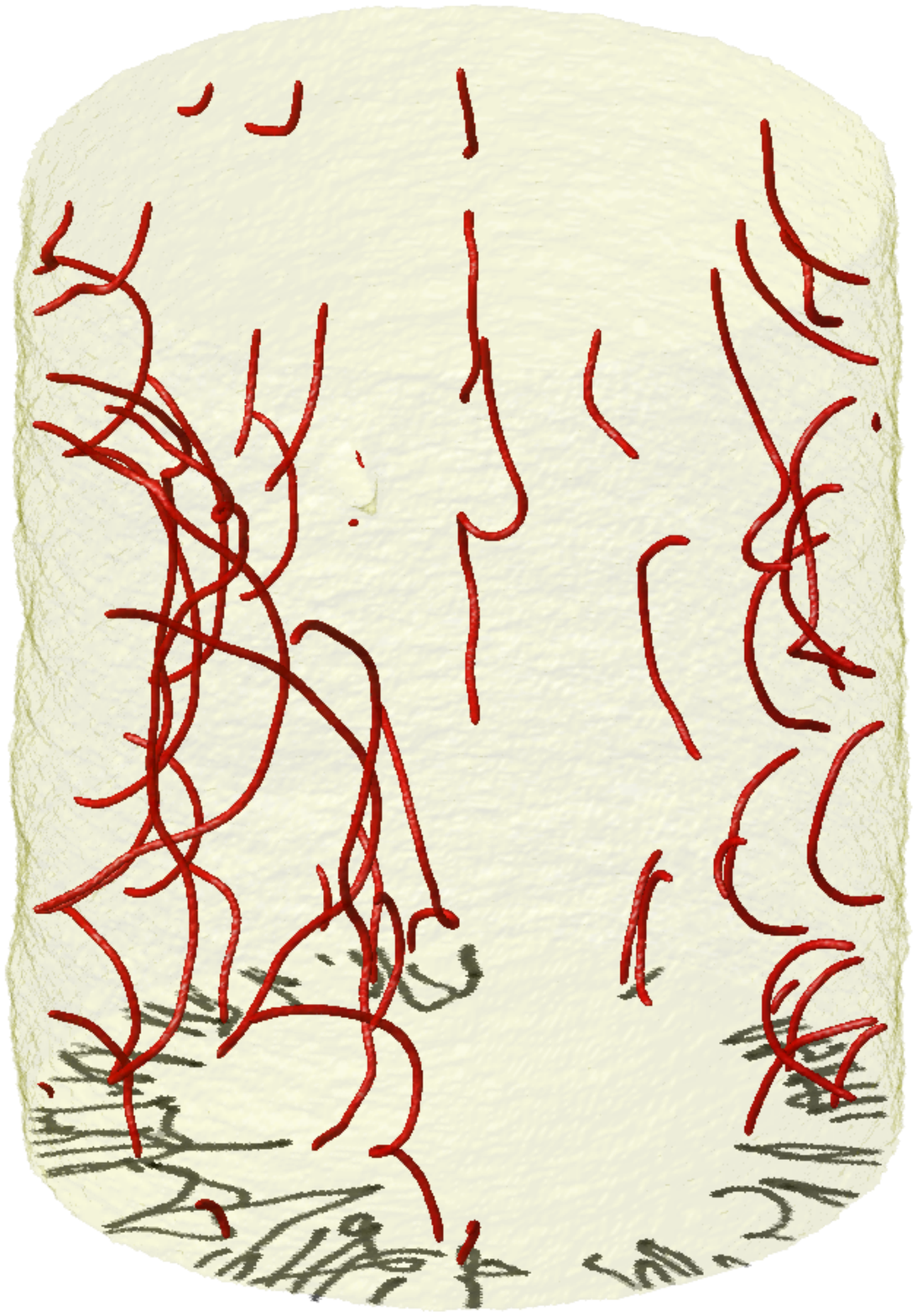}
		\includegraphics[width = 0.19\textwidth]{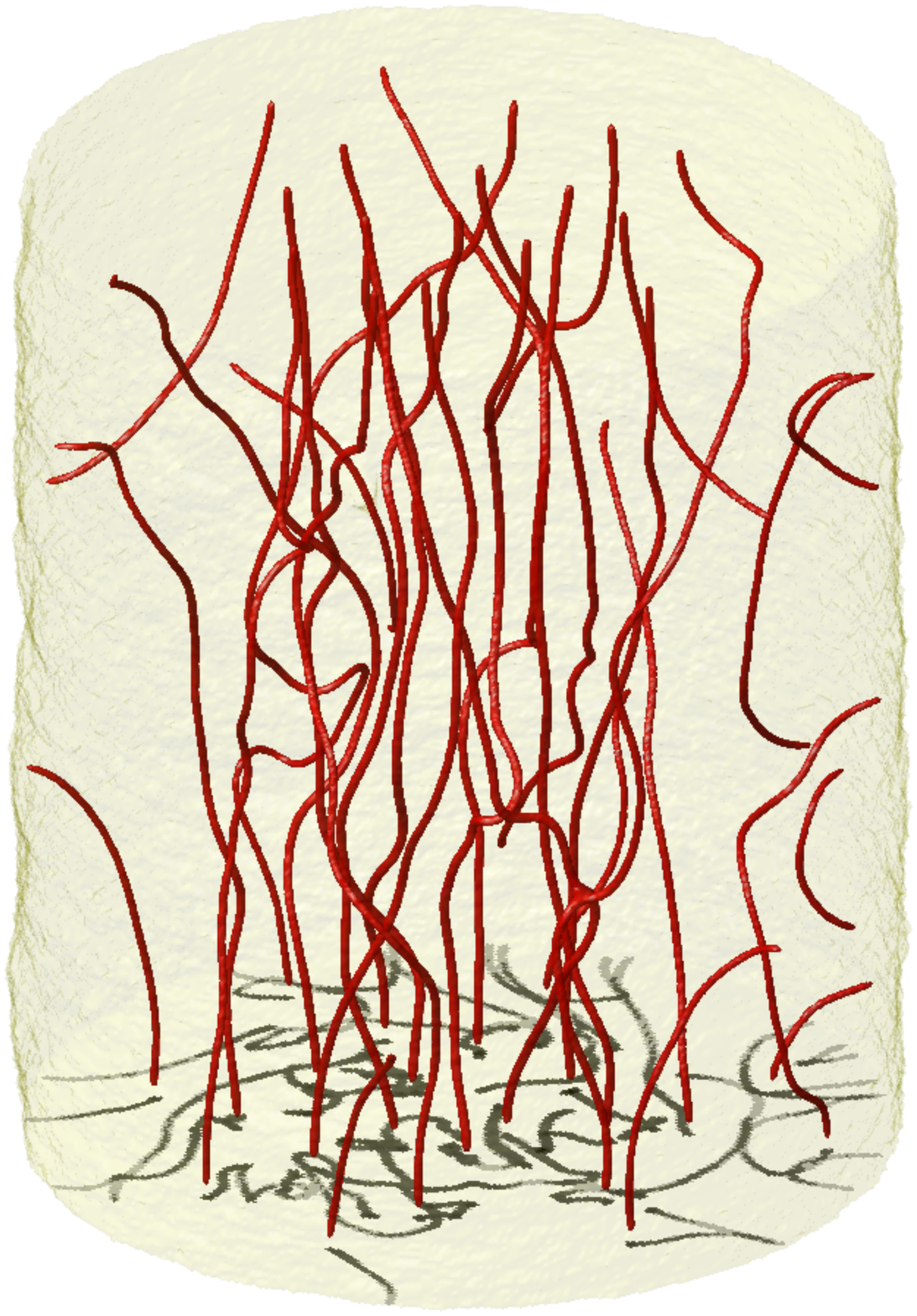}	
\caption{Early-time dynamics during the spin-up of the fluid in the 
presence of a single strong protuberance added to the rough
cylindrical boundary. The snapshots,  taken at 
$t/\tau=26$, $50$, $100$, $200$ and $500$, are presented in the same way 
as Fig.~\ref{fig2}. }
\label{fig7}
\end{figure*}

First we consider the effect of a single strong imperfection  in the
form of a protuberance on the cylindrical wall. The question is
whether, by enhancing vortex nucleation, the protuberance 
can induce a turbulent boundary layer.
The protuberance is numerically created by adding a Gaussian-shaped 
potential to the existing (small-scale) roughness potential. 
Equation~\eqref{eq:rbucket} is replaced by

\begin{equation}
r(z,\phi)=R[1-a (f(z,\theta) + G f_G(z,\theta))],
\label{eq:rbucket2}
\end{equation}

\noindent
where $G=2$ and $f_G(z,\theta)$ is a Gaussian-shape function taking values
from $0$ to $1$ and rms width $4 \xi$. The approximate height of the 
strong protuberance 
in the simulation which we present is $10 \xi$, as also visible
in Fig.~\ref{fig6}(f).

\begin{figure}
\begin{center}
\includegraphics[width = 0.45\textwidth]{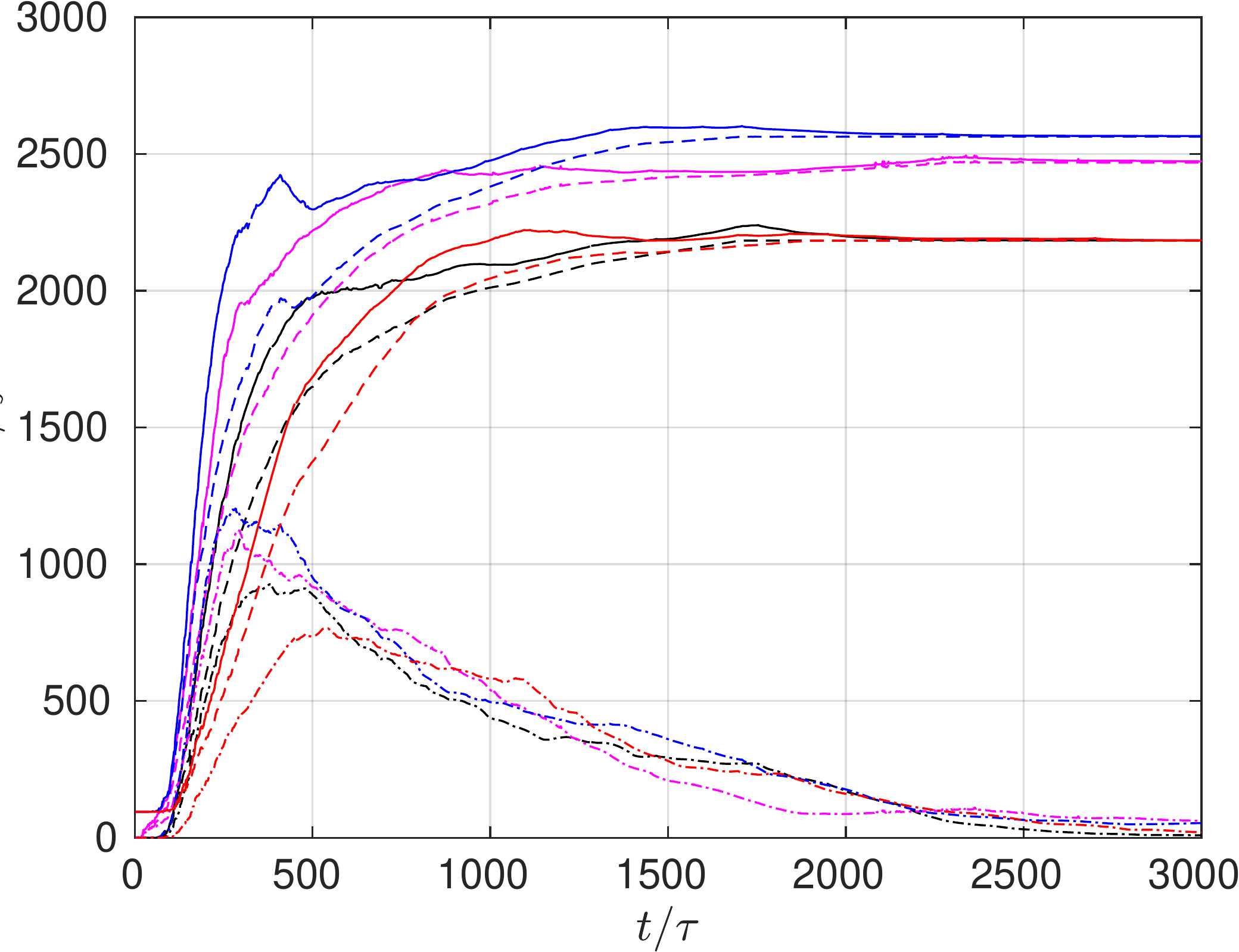}
\caption{
Time evolution of $\Lambda$ (solid line), $\Lambda_z$ (dashed line), 
$\Lambda_{xy}$ (dot-dashed line). The colours correspond to the 
simulations with default parameters $\Omega=0.02 \tau^{-1}$,
$a=0.1 \xi$ (black),
the added surface protuberance (magenta),
the added remanent negative vortex (blue), and
the added remanent positive vortex (red), respectively.
}
\label{fig10}
\end{center}
\end{figure}

\noindent
Snapshots taken during the time evolution for $\Omega=0.02~\tau^{-1}$
and $a=0.1$ are shown in Fig.~\ref{fig7}; a movie can be viewed 
in Supplementary Material  \cite{movie_gaussian}.
The protuberance catalyses the local nucleation of vortices 
at early times: large vortex loops 
(of the same size order as the protuberance) are rapidly generated
[Fig.~\ref{fig7}(a)], leading to a downstream trail of 
loops [Fig.~\ref{fig7}(b, c)], in addition to the slower 
nucleation of U-vortices from the rough bucket wall. 
The vortex configuration becomes
clearly anisotropic near the bucket edge [Fig.~\ref{fig7}(d)].  
However, once the vortices fill the bulk [Fig.~\ref{fig7}(e)],
memory of this effect is lost, and the subsequent evolution is
very similar to the evolution without the strong protuberance.
In fact, the final vortex lattice is not significantly different from the
lattices considered in Section~\ref{section3}, as shown in Fig.~\ref{fig6}(f). Figure~\ref{fig10} shows the time evolution of $\Lambda$,
$\Lambda_z$ and $\Lambda_{xy}$ in the presence of the protuberance (magenta lines) and its absence (black lines).  This confirms that the protuberance accelerates the generation of vortex line length at early times, but that its effect becomes washed out at later times.

\subsection{Effect of remanent vortices}

Secondly, we consider the effect of remanent vortex lines.
In experiments with liquid helium, it is believed that so-called 
`remanent vortices' may be present in the fluid, created via the 
Kibble-Zurek mechanism when cooling the helium sample through the 
superfluid transition to the final experimental temperature.  
The presence of remanent vortices may modify the vortex nucleation and the
formation of the vortex lattice when the sample is rotated.
To explore this idea, we have repeated the simulations imposing
a suitable phase profile to add a vortex to the initial state during the 
imaginary-time propagation.  
For simplicity we position the remanent vortex along the $z$-axis 
of rotation. 

The evolution of the superfluid with the standard rough cylindrical
wall and a ``positive" remanent vortex, that is, one whose circulation is oriented in the same direction of the
bucket's rotation is shown through Fig.~\ref{fig8} 
and the movie in the Supplementary Material \cite{movie_pos_remnant}. 
Compared to Section~\ref{section3},
the only significant modification is a dampening of the initial injection 
of U-vortices; the effect is visible by eye when comparing like-time 
snapshots [Fig. \ref{fig2}(b) and Fig.~\ref{fig8}(a)].  
The remanent vortex acts in the same direction as the rotating container: 
it reduces the relative speed between the bucket's wall and the superfluid,
and remains largely undisturbed at early times 
[Fig.~\ref{fig8}(a)] until the 
U-vortices that are nucleated fill the bulk and interact 
with it [Fig.~\ref{fig8}(b)]; at this point
the remanent vortex becomes subsumed within the other like-signed 
vortices [Fig.~\ref{fig8}(c)], and the 
subsequent relaxation of the vortex configuration into a vortex
lattice largely proceeds as if there was not any remanent vortex initially. Confirming this, we see that in Figure~\ref{fig10} that the presence of the positive vortex (red lines) depletes the generation of vortex line length at early times, but this recovers at later times such that the system reaches the same line length as in the absence of any remanent vortices (black lines).

\begin{figure*}
\hspace{-2.2cm} (a) \hspace{3.7cm} (b) \hspace{3.9cm} (c)\\
\includegraphics[width = 0.22\textwidth]{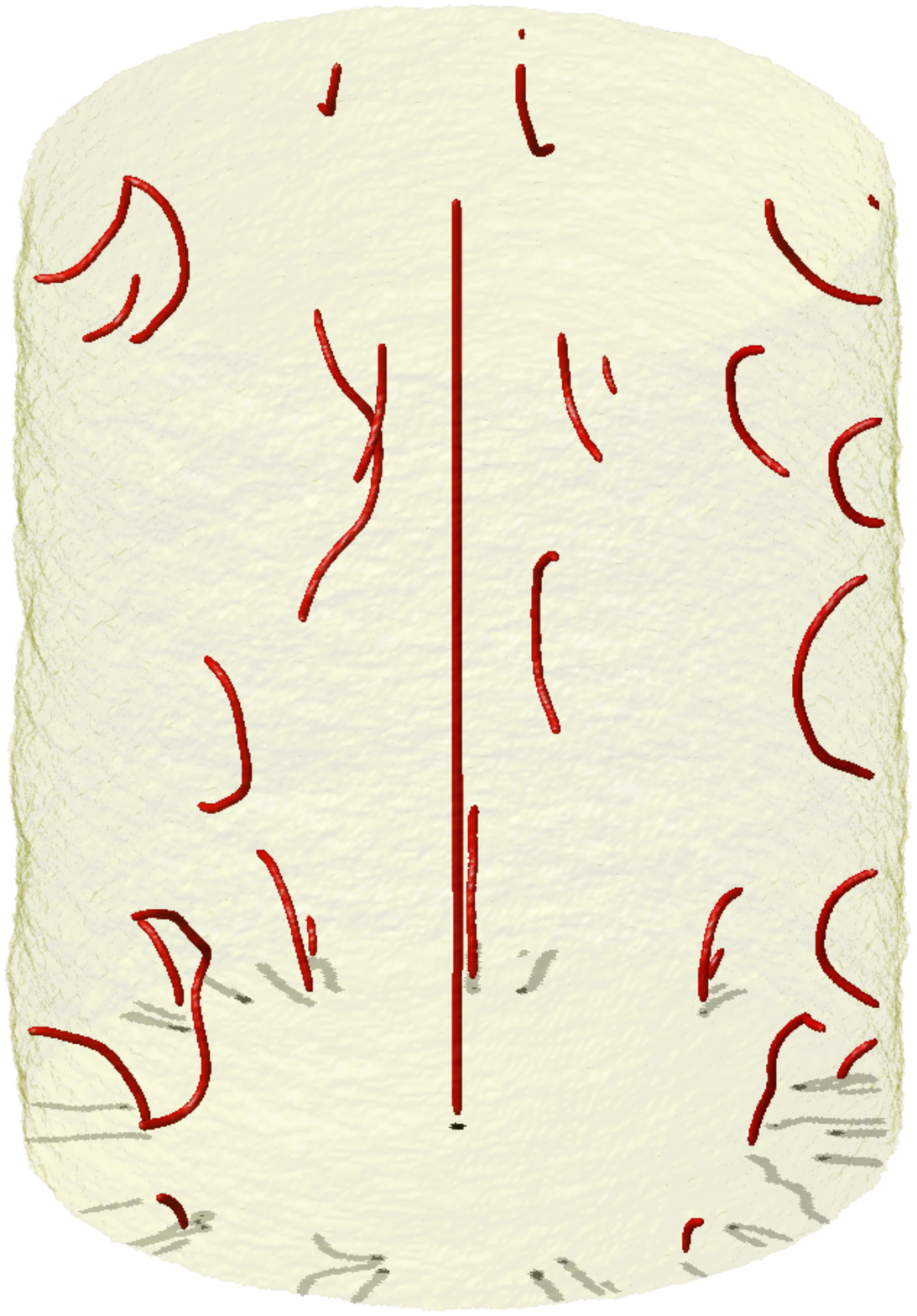}
\includegraphics[width = 0.22\textwidth]{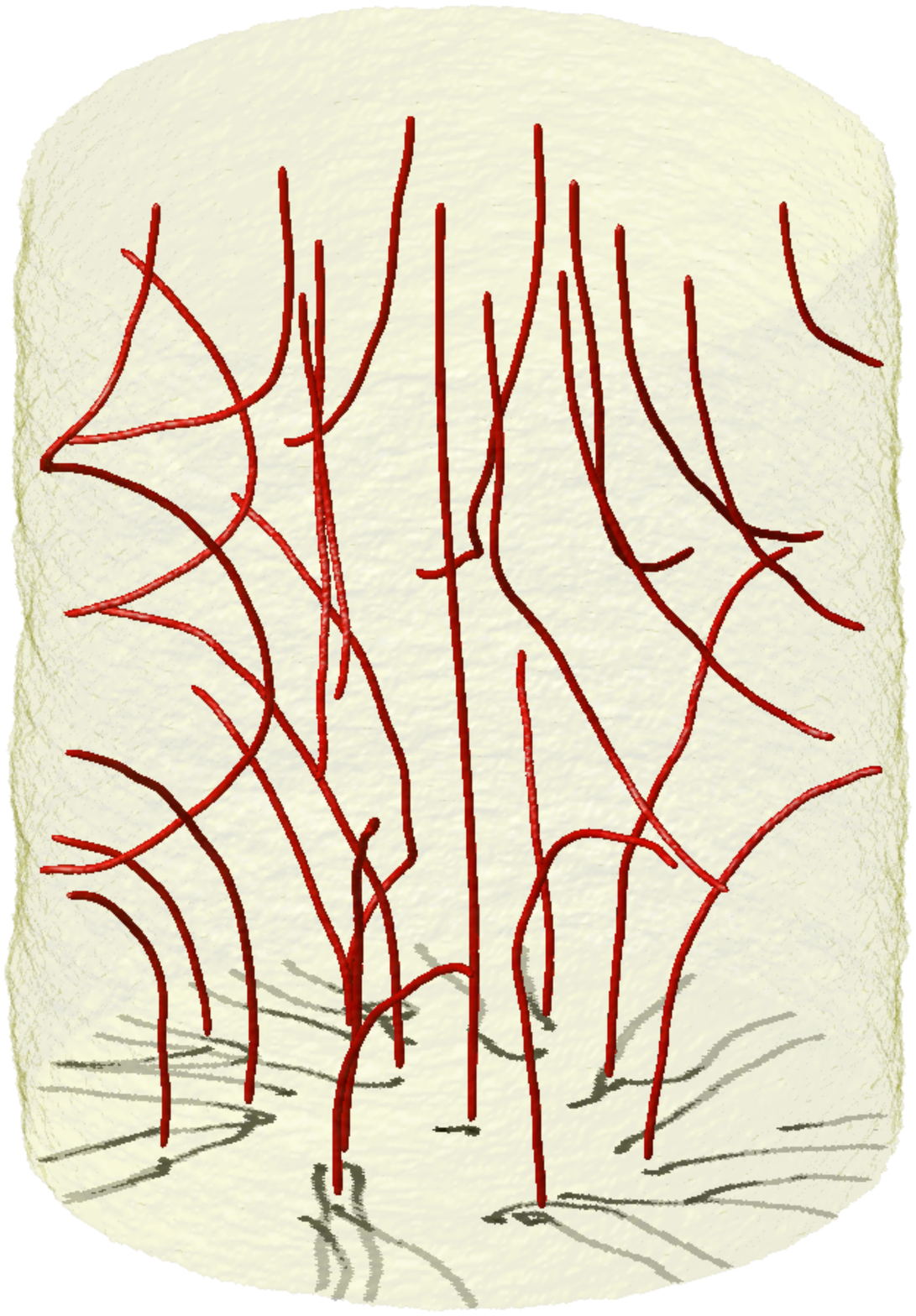}
\includegraphics[width = 0.22\textwidth]{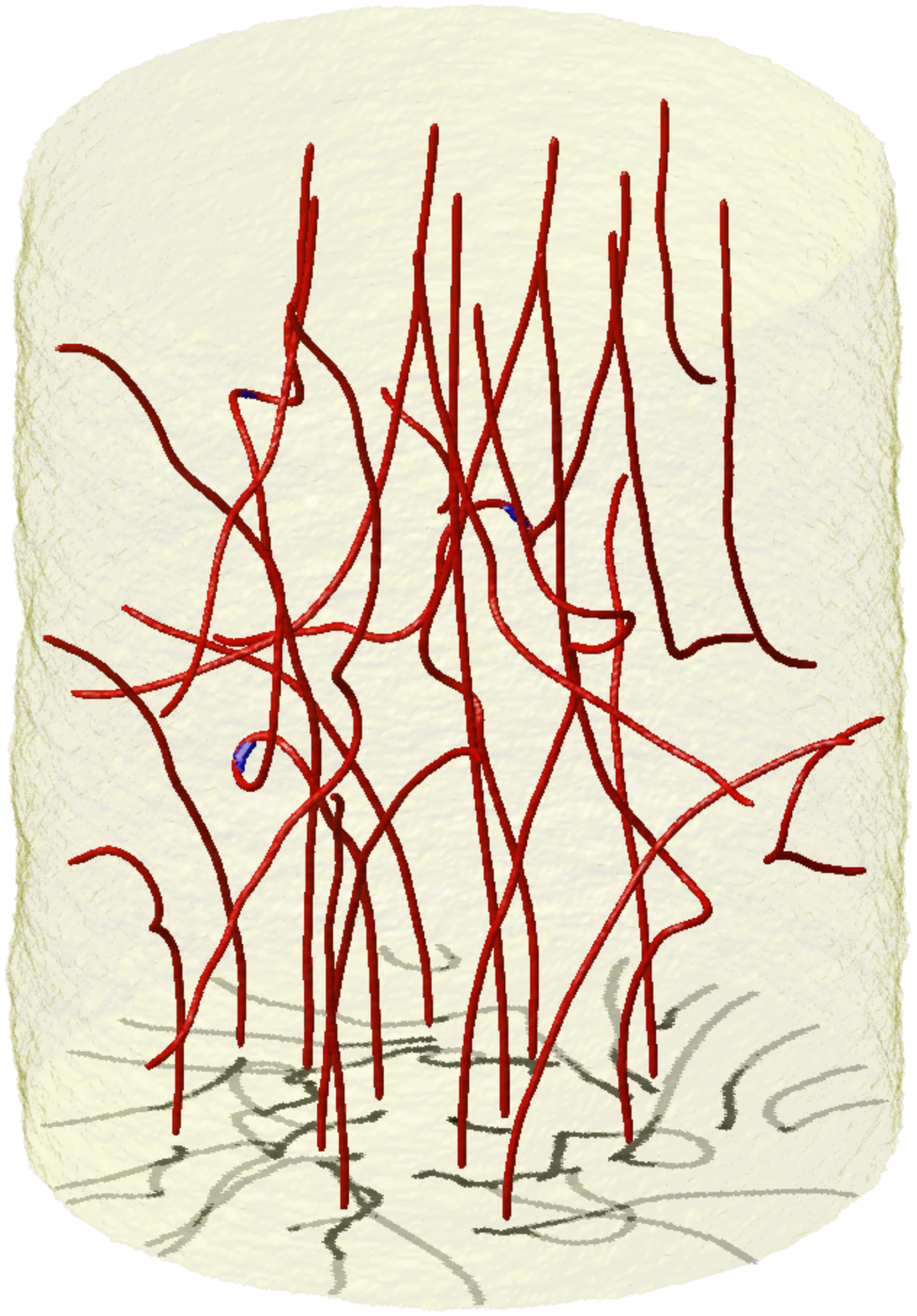}
\caption{Spin-up of the superfluid in the presence of a positively-charged 
remanent vortex.  Snapshots are taken at $t/\tau=200$, $400$ and $500$, 
and are presented in the same way as Fig.~\ref{fig2}.}
\label{fig8}
\end{figure*}

\begin{figure*}
\hspace{0.2cm} (a) \hspace{3.7cm} (b) \hspace{3.2cm} (c)
\hspace{0.2cm} (d) \hspace{3.7cm} (e) \hspace{3.2cm} 
\includegraphics[width = 0.19\textwidth]{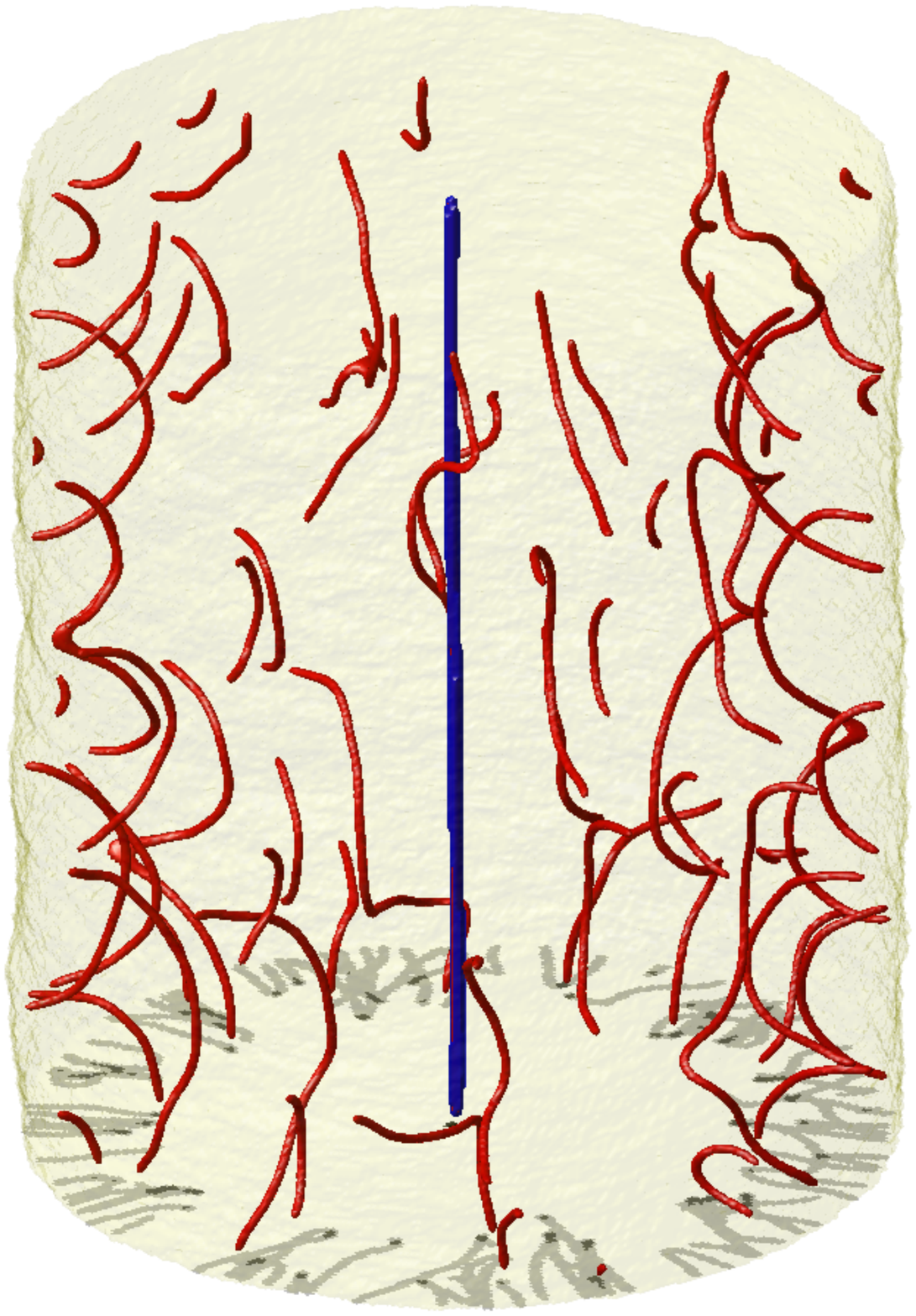}
\includegraphics[width = 0.19\textwidth]{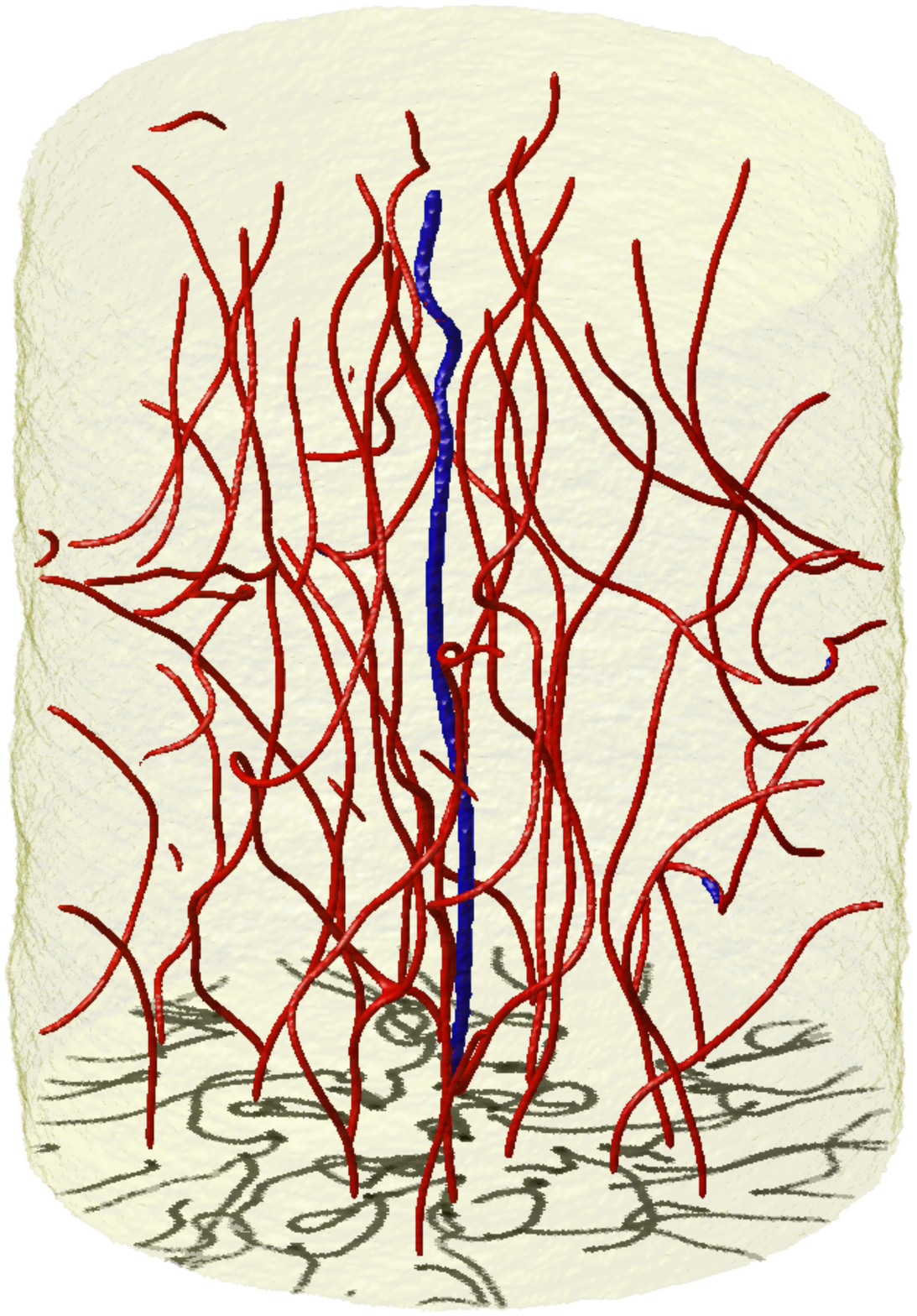}
%\hspace{0.2cm} (c) \hspace{3.7cm} (d) \hspace{3.2cm} 
\includegraphics[width = 0.19\textwidth]{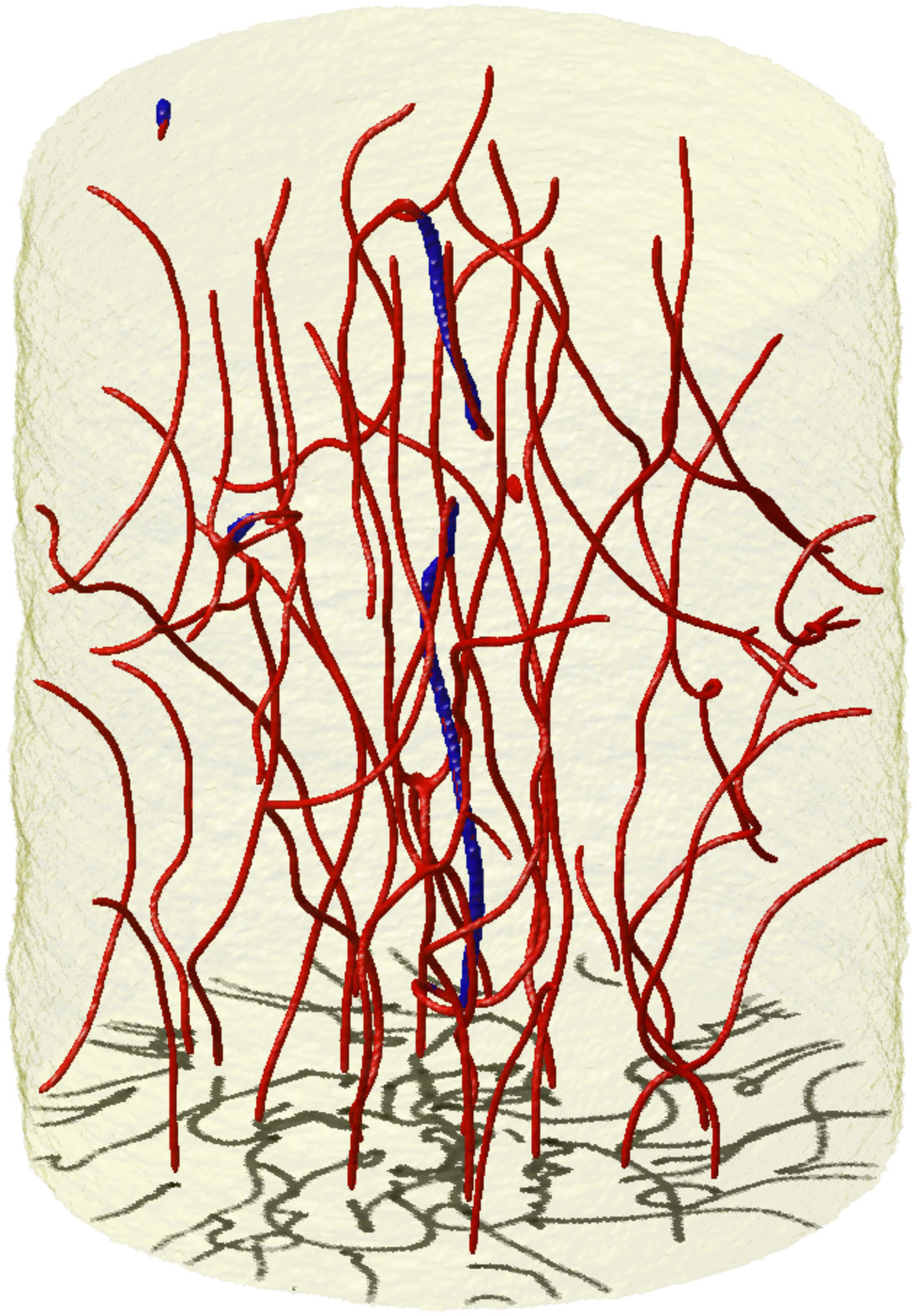}
\includegraphics[width = 0.19\textwidth]{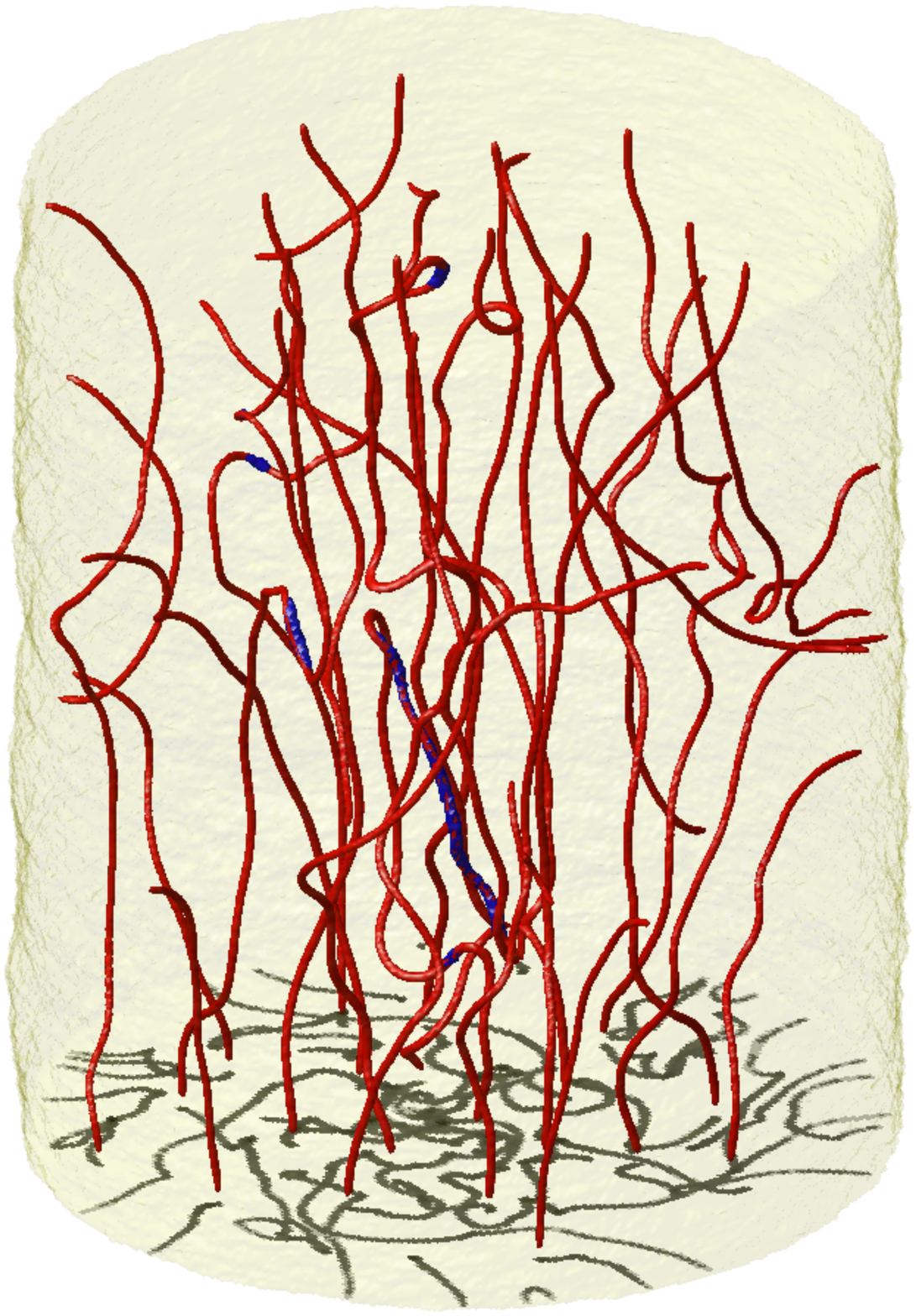}
%\hspace{0.2cm} (e) \\
\includegraphics[width = 0.19\textwidth]{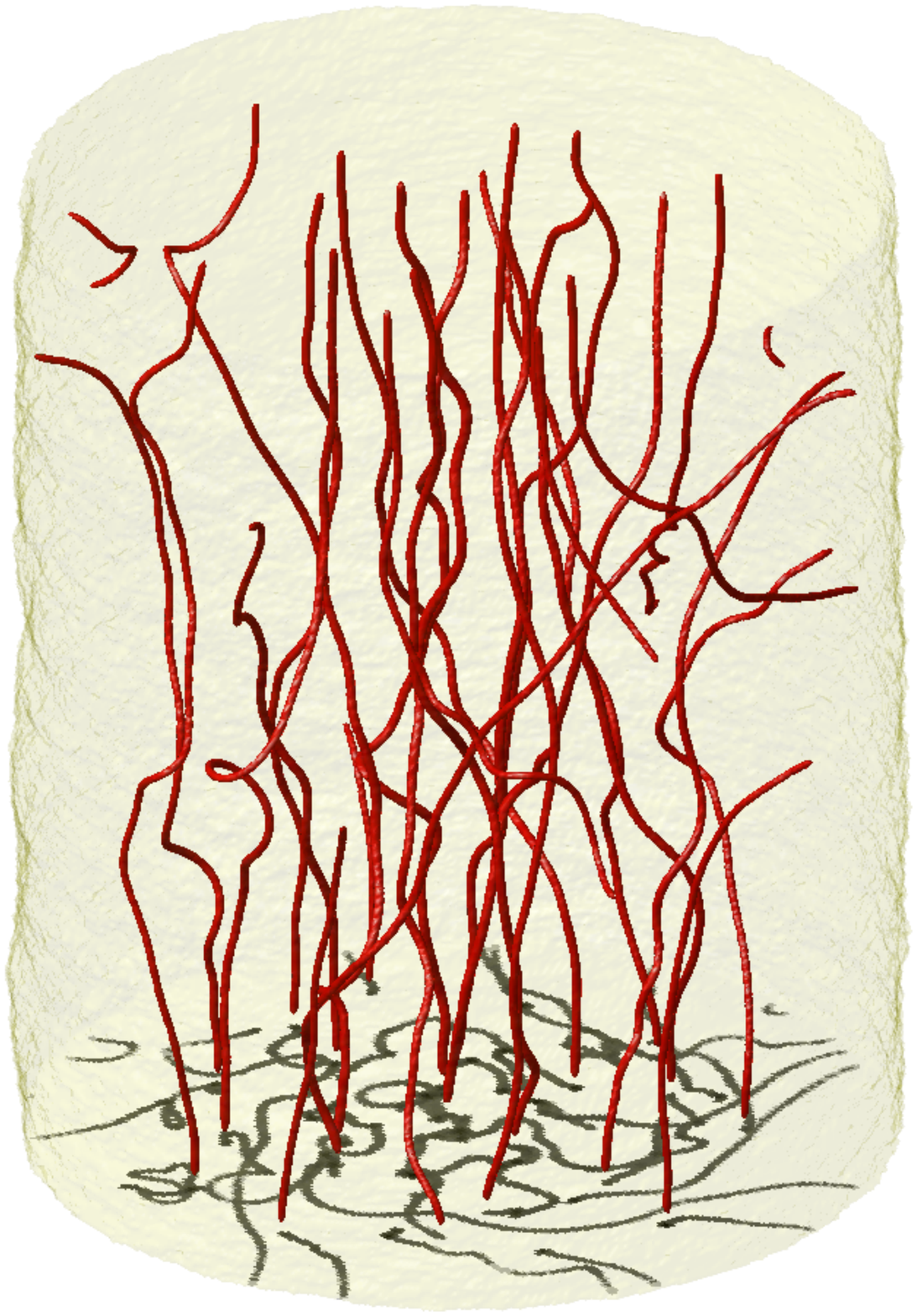}
\caption{Spin-up of the superfluid in the presence of a negatively-charged 
remanent vortex.  
Snapshots are taken at $t/\tau=200$ (a),  $360$ (b), $386$ (c), $420$ (d)
and $500$ (e). The images are presented in the same way as Fig.~\ref{fig2}.  
Vortices with negative circulation are coloured blue.}
\label{fig9}
\end{figure*}

If the remanent vortex is oriented in the direction opposite to the
rotation of the bucket, i.e. a negative vortex, the evolution proceeds differently, as seen in Fig.~\ref{fig9} and the movie in Supplementary Material
\cite{movie_neg_remnant}.  
The remanent vortex enhances the nucleation of U-vortices
from the boundary, as evident from comparing 
Fig.~\ref{fig2}(b) and Fig.~\ref{fig9}(a).  
This effect is caused by the counter-flow induced by the remanent
vortex, which increases the relative speed of the fluid over the rough 
boundary. 
Once the other vortices drift close to the remanent vortex, the 
remanent vortex becomes excited by their interaction
[Fig.~\ref{fig9}(b)].  
A series of vortex reconnections break up the remanent vortex, 
forming progressively smaller vortex loops 
[Fig.~\ref{fig9}(c,d)].  This leads to the rapid 
removal of vorticity of the `wrong' sign from the fluid
 [Fig.~\ref{fig9}(e)].  
Hereafter the fluid evolves in a similar manner to when the remanent vortex in absent [Section~\ref{section3}], albeit with a slightly higher final vortex line length [Fig.~\ref{fig10}].

\subsection{2D case}

Finally, we have also performed the corresponding 2D simulations of the spin-up of a 2D superfluid within a rough circular boundary; the boundary is taken from the central slice of the 3D rough bucket.  A movie showing the typical dynamics
is available in the Supplementary Material \cite{movie_2d_large_bucket}.
The 2D geometry
allows calculations of much larger buckets, up to $R=200 \xi$ with
a $1024^2$ numerical grid.  We observe the same qualitative behaviour as in
3D in smaller buckets, albeit with many more vortices and without 3D effects
such as vortex reconnections. Collisions of vortices of the opposite
circulation result in the annihilation of the vortices and the emission of
sound pulses \cite{Kwon2014,Stagg2015,Groszek2016}.
In general, we find that, in 2D, the timescales of injection, diffusion and 
lattice crystallisation are faster than in 3D.   
A particular feature that we see in the early-time dynamics of
the 2D simulations is the nucleation of vortices with both positive 
and negative circulation  (i.e. with circulation which is inconsistent
with the imposed rotation).  We notice that some negative vortices 
originate from localised rarefaction pulses generated 
from the rough boundary when the bucket is set into rotation. 
We associate these pulses with Jones-Roberts solitons 
\cite{Jones1982,Tsuchiya2008}, which are low energy/momentum 
solutions of the 2D GPE.  
At higher energy/momentum, these solutions become pairs of vortices 
of opposite sign (also called vortex dipoles in the literature).
The conversion of Jones-Roberts solitons into vortex dipoles occurs
if the pulse gains energy from the large positive vortex cluster
which starts forming in the centre of the bucket. Occasionally,
the vortices which are parts of a dipole separate
and mix with the rest of the vortices.  Over time, the vortices 
of negative circulation are lost from the system, either 
colliding (hence annihilating) with positive vortices within the bulk,
or by exiting the fluid at the bucket's boundary (effectively
annihilating with their images).  
\vspace{0.3cm}

\section{Conclusions}
\label{section5}

In conclusion, we have employed simulations of the Gross-Pitaevskii equation to study the spin-up of a superfluid in a rotating bucket featuring microscopically rough walls.   Within this model, we see several key stages of the dynamics.  Firstly, vortices are nucleated at the boundary by the flow over the rough features, typically in the form of small U-shaped vortex lines.  Secondly, these U-shaped vortices interact strongly and reconnect, creating
a transient turbulent state.  This becomes increasingly polarised by the imposed rotation until the vortex configuration consists of vortices of the correct orientation extending from the top to the bottom of
the bucket.  Finally, the vortex lines slowly straighten and arrange themselves in the expected final vortex lattice configuration. Our results highlight the importance of vortex 
reconnections \cite{Galantucci2019}: it is
generally assumed that vortex reconnections are important in turbulence,
but here we have seen that reconnections are essential to create, starting
from potential flow, something as simple as solid body rotation
(the vortex lattice). 
The addition
of a single large protuberance or one additional remanent vortex line does not
change the dynamics significantly, only speeding up or slowing down the injection of vorticity.  Moreover, analogous dynamics  arise in the 2D limit. 

We reiterate that the GPE is not a quantitatvely accurate model of superfluid helium and these results should be interpreted qualitatively only.   For example, the role of friction is introduced into the GPE through a widely-used phenomenological dissipation term; however, a more accurate physical model of this stage of the dynamics would be provided by the VFM.  Also, a distinctive physical property of superfluid helium is its strong non-local interactions.  This, for instance, supports a roton minimum in its excitation spectrum.  While this is absent from the GPE model we have employed, it can be introduced through an additional non-local term \cite{Berloff2014,Reneuve2018}.  It would be interesting to see if this causes any significant departures from the dynamics we have reported.

\section*{Acknowledgements}

N.P., L. G. and C.F.B. acknowledge support by the Engineering and Physical 
Sciences Research Council (Grant No. EP/R005192/1).

\end{document}